\documentclass[twocolumn,floats,floatfix,showpacs,amssymb,prd,superscriptaddress,nofootinbib,eqsecnum]{revtex4-1}
\usepackage{graphicx, epsfig, amssymb} 
\usepackage{amsmath, amsfonts,amsthm,latexsym}
\usepackage{bm} 

\usepackage[linktocpage]{hyperref}
\usepackage[caption=false]{subfig}
\usepackage[usenames]{color}

\def\be{\begin{equation}}
\def\ee{\end{equation}}
\def\beq{\begin{eqnarray}}
\def\eeq{\end{eqnarray}}

\begin{document}

\title{\large Numerical relativity for $D$ dimensional
       space-times: \\ Head-on collisions of black holes and gravitational wave extraction}

\author{Helvi Witek}\email{helvi.witek@ist.utl.pt}
\affiliation{
  Centro Multidisciplinar de Astrof\'\i sica --- CENTRA,
  Departamento de F\'\i sica, Instituto Superior T\'ecnico --- IST \\
  Universidade T\'ecnica de Lisboa - UTL,
  Av. Rovisco Pais 1, 1049-001 Lisboa, Portugal  
}

\author{Miguel Zilh\~ao}\email{mzilhao@fc.up.pt}
\affiliation{
  Centro de F\'\i sica do Porto --- CFP,  
  Departamento de F\'\i sica e Astronomia,
  Faculdade de Ci\^encias da Universidade do Porto --- FCUP \\
  Rua do Campo Alegre, 4169-007 Porto, Portugal
}

\author{Leonardo Gualtieri}\email{leonardo.gualtieri@roma1.infn.it}
\affiliation{
  Dipartimento di Fisica, Universit\`a di Roma
  ``Sapienza'' \& Sezione \\
  INFN Roma1, P.A. Moro 5, 00185, Roma, Italy
}

\author{Vitor Cardoso}\email{vitor.cardoso@ist.utl.pt}
\affiliation{
  Centro Multidisciplinar de Astrof\'\i sica --- CENTRA,
  Departamento de F\'\i sica, Instituto Superior T\'ecnico --- IST \\
  Universidade T\'ecnica de Lisboa - UTL,
  Av. Rovisco Pais 1, 1049-001 Lisboa, Portugal 
}
\affiliation{
  Department of Physics and Astronomy, The University of Mississippi,
  University, MS 38677-1848, USA
}

\author{Carlos Herdeiro}\email{herdeiro@ua.pt}
\affiliation{Departamento de F\'\i sica da Universidade de Aveiro, 
Campus de Santiago, 3810-183 Aveiro, Portugal.}
\affiliation{Centro de F\'\i sica do Porto --- CFP,
Departamento de F\'\i sica e Astronomia,
Faculdade de Ci\^encias da Universidade do Porto --- FCUP,
Rua do Campo Alegre, 4169-007 Porto, Portugal.}

\author{Andrea Nerozzi}\email{andrea.nerozzi@ist.utl.pt}
\affiliation{
  Centro Multidisciplinar de Astrof\'\i sica --- CENTRA,
  Departamento de F\'\i sica, Instituto Superior T\'ecnico --- IST \\
  Universidade T\'ecnica de Lisboa - UTL,
  Av. Rovisco Pais 1, 1049-001 Lisboa, Portugal 
}

\author{Ulrich Sperhake}\email{sperhake@tapir.caltech.edu}
\affiliation{
  Department of Physics and Astronomy, The University of Mississippi,
  University, MS 38677-1848, USA
}
\affiliation{
  Institut de Ci\`encies de l'Espai (CSIC-IEEC),
  Facultat de Ci\`encies, Campus UAB,
  Torre C5 parells, E-08193 Bellaterra, Spain
}
\affiliation{
  California Institute of Technology,  
  Pasadena, CA 91125, USA
}


\begin{abstract}

  Higher dimensional black holes play an exciting role in fundamental physics, 
  such as high energy physics.
  In this paper, we use the formalism and numerical code reported
  in~\cite{Zilhao:2010sr} 
  to study the head-on collision of two black holes. For
  this purpose we provide a detailed treatment of gravitational wave extraction
  in generic $D$~dimensional space-times, which uses the Kodama-Ishibashi
  formalism. For the first time, we present the results of numerical simulations
  of the head-on collision in five space-time dimensions, together with the
  relevant physical quantities. We show that the total radiated energy, when two
  black holes collide from rest at infinity, is approximately $(0.089\pm
  0.006)\%$ of the centre of mass energy, slightly larger than the $0.055\%$
  obtained in the four dimensional case, and that the ringdown signal at late
  time is in very good agreement with perturbative calculations.

\end{abstract}


\maketitle


\section{Introduction}
Black objects in higher dimensional space-times have a remarkably
richer structure than their four dimensional counterparts. They appear in a variety
of configurations (e.g.\ black holes, black branes, black rings,
black Saturns), and display complex stability phase diagrams 
(see \cite{Emparan:2008eg} for an overview). 
They might also play a key role in high energy physics: 
%
High energy physics scenarios, such as the gauge-gravity
duality \cite{Maldacena:1997re} or TeV gravity models
\cite{Antoniadis:1990ew,ArkaniHamed:1998rs,Antoniadis:1998ig,
Randall:1999ee,Randall:1999vf},
suggest that dynamical processes involving higher dimensional black holes
(BHs) may be relevant for understanding the physics under experimental
scrutiny at particle colliders, such as the Large Hadron Collider (LHC) or
the Relativistic Heavy Ion Collider (RHIC). These processes can be quite
violent and highly nonlinear, as in the case of BH collisions. 
Numerical relativity, which solves Einstein's equations 
on supercomputers, is therefore the only 
available tool for high-precision studies of such BH systems. 
Fortunately, the field of numerical relativity has matured considerably over the last
five years (see Refs.~\cite{Pretorius:2007nq,Hinder:2010vn} for reviews),
and its techniques can now be extended to a much wider class of space-times.
Space-times of generic dimensionality or with more general asymptotics,
feature most prominently among such generalizations.
Recent applications to the study of higher dimensional BH instabilities may be found in
Refs.~\cite{Shibata:2009ad,Shibata:2010wz}, using the formalism developed
in Ref.~\cite{Yoshino:2009xp}; an application to the study of $AdS$-like
asymptotics may be found in Ref.~\cite{Witek:2010qc}. Further applications of numerical relativity to more general types of space-times
have been discussed in Ref.~\cite{Zilhao:2010sr}, hereafter denoted as Paper I, wherein
we have started a long-term effort to evolve BH space-times in higher dimensions numerically, 
and developed a framework to perform numerical simulations of $D$~dimensional 
space-times with an $SO(D-2)$ isometry group (for $D\ge5$) or $SO(D-3)$ 
(for $D\ge6$).

One scenario in which BH collisions play a very well defined role is
that of TeV-scale gravity, i.e. scenarios in which the fundamental
Planck scale is of the order of the TeV. The beginning
of the scientific runs at the LHC makes accurate
theoretical modelling of the experimental signatures of this scenario
for LHC collisions very timely. In this scenario, for centre of
mass energies well above the TeV threshold,
recall that LHC collisions will reach 14 TeV, parton-parton collision will be dominated
by the gravitational interaction, and should be well described
by any classical gravitational objects with the same gravitational
energy. For modelling simplicity, it is convenient to
choose these objects to be BHs
\cite{Banks:1999gd,Giddings:2001bu,Dimopoulos:2001hw,Feng:2001ib,Ahn:2003qn,
  Ahn:2002mj,Chamblin:2004zg,Cardoso:2004zi,Cavaglia:2002si,Kanti:2004nr,
  Kanti:2008eq,Solodukhin:2002ui,Hsu:2002bd}. Due to 
the dominance of the gravitational interaction,
we may further neglect the electric charge of the holes;
charge dependant effects should give sub-leading corrections to the
relevant observables. For sufficiently small impact parameters,
these trans-Planckian collisions are expected to form a BH
\cite{Giddings:2001bu,Dimopoulos:2001hw}, as follows from Thorne's hoop
conjecture \cite{Thorne:1972ji}
which recently has received 
support from the numerical work of Choptuik and Pretorius \cite{Choptuik:2009ww}.
Therefore, there is substantial evidence that modelling the individual partons as BHs is
not biasing the final result of the parton scattering process towards BH
formation. Indeed, this is the idea that, above the fundamental Planck
scale, \textit{matter does not matter}; it only matters the gravitational
energy that each parton carries.

After formation, the BH should then decay via Hawking evaporation. In
order to filter experimental data, this process has been modelled by
dedicated Monte Carlo event generators, such as \textsc{Truenoir},
\textsc{Catfish}, \textsc{Charybdis2} or \textsc{Blackmax}
\cite{Dimopoulos:2001hw,Cavaglia:2006uk,Frost:2009cf,Dai:2007ki,Dai:2009by}.
The latter two are being used by the ATLAS experiment at the LHC. These
generators clearly exhibit the very distinct experimental signatures
of BH evaporation, including higher multiplicity of jets and larger transverse
momentum than those produced by any standard model process \cite{Aad:2009wy}.
The event generators also model the BH production phase from the parton-parton scattering,
for which they need as an input the threshold impact parameter
for BH formation and the energy lost in gravitational radiation
during the parton-parton collision. At the moment, the best
estimates for these quantities are based on trapped surface methods
\cite{Yoshino:2005hi}. Accurate results can, however, be obtained from
full-blown numerical simulations, as has already been seen in four dimensions
\cite{Sperhake:2008ga,Shibata:2008rq,Sperhake:2009jz}. Such results
will be instrumental in a more accurate phenomenological modelling of
BH production/evaporation in particle colliders. Observe that, even
if no evidence for BH formation/evaporation is found at the LHC, such
accurate modelling will matter for setting precise lower bounds on the
fundamental Planck scale.

In this paper we present the first fully nonlinear treatment of a
head-on collision of BHs in a higher dimensional space-time, together
with an analysis of the relevant physical quantities. This is achieved
by solving the corresponding Einstein equations numerically, for which
we use the formalism and code reported in Paper I.
Once given the numerically constructed space-time, 
one is still left with the question of extracting physically meaningful 
quantities, such as the energy and linear and angular
momentum carried away by the gravitational radiation.

In four space-time dimensions, two distinct formalisms
have been developed to extract the physical information (see
e.g. Ref.~\cite{Alcubierre:2008} for a review of both formalisms). One
is based on the Regge-Wheeler-Zerilli perturbation theory for the
Schwarzschild BH \cite{Regge:1957td,Zerilli:1970se}; the other
is based on the Newman-Penrose formalism \cite{Newman:1961qr},
which was used by Teukolsky to study perturbations of
algebraically special space-times \cite{Teukolsky:1973ha},
a class which includes the Kerr BH. A higher dimensional
generalization of the Newman-Penrose formalism has been developed
in Refs.~\cite{Coley:2004jv,Milson:2004jx,Pravda:2004ka,Ortaggio:2007eg}.
Unfortunately, the condition of being algebraically special does not seem
to be as powerful for the study of exact solutions or their perturbations
in higher dimensions as it was in four dimensions. 
For instance, the Goldberg-Sachs theorem is not valid any longer in higher dimensions
\cite{Pravda:2004ka,Ortaggio:2007eg}. For the study we present herein,
however, it suffices to use the higher dimensional generalisation
of the Regge-Wheeler-Zerilli formalism, 
because the final result of a head-on collision of two $D$~dimensional, 
nonspinning BHs approaches, at late times, a $D$~dimensional Schwarzschild, 
i.e. Tangherlini \cite{Tangherlini:1963bw} BH. Fortunately, the perturbation theory
of the latter BH has been fully developed, in arbitrary dimensions,
by Kodama and Ishibashi \cite{Kodama:2003jz}. Our remaining task is to
obtain the relevant gauge-invariant quantities from our numerical data,
a procedure that we shall describe in detail in this paper.

After implementing this wave-extraction formalism we apply it to study
the head-on collision of two BHs in four and five dimensional
space-times.  In four dimensions we recover previous results in
the literature, and 
we perform a number of tests on the numerical coordinate system, to ensure it is
appropriate for the wave extraction formalism.  We estimate that around
$0.055\%$ of the centre of mass energy is radiated when two BHs, at rest at infinity, collide.
This result is in good agreement with those reported in the literature \cite{Sperhake:2006cy}.
The five dimensional results are entirely new. We show that the kinematics of 
the BHs before the merging follow, to a good precision, the Newtonian prediction. 
We estimate that around $0.089\%$ of the centre of mass energy is radiated as
gravitational waves, when two BHs collide from rest at infinity, and present the associated waveforms.
We stress that these results refer to a fully nonlinear
evolution of Einstein's field equations.

This paper is organized as follows. In Section \ref{gwaves} we
apply the Kodama-Ishibashi (KI) formalism to the space-times considered
in Paper I.  By assuming that the numerical space-time is a small
deviation from the Tangherlini solution, one is able to relate
(see relations~(\ref{metricrelation})) the numerical metric
to the KI metric perturbations
and to compute gauge-invariant quantities using Eqs.~(\ref{gi1}),
(\ref{gi2}). These are then used to construct a master function $\Phi$ (see Eq.~(\ref{eq:KIwavefunction})) from which all 
relevant information about the radiation can be computed.
In Sections~\ref{numres}, \ref{sec:5D} we present results obtained from the evolution of
Brill-Lindquist initial data in $D=4,~5$ respectively, that represents
the collision of two equal-mass, nonspinning BHs which are initially at rest.
In order to calibrate the accuracy of the wave extraction formalism, 
we perform a number of tests, including tests on the numerical coordinates
themselves. We compute the time derivative of the master function $\Phi$, energy fluxes and
total energy radiated. We give some final remarks in Sec.~\ref{final} and 
discuss future steps in this research program. Two appendices 
cover technical details on the coordinate transformation between the numerical and wave extraction
frames (Appendix~\ref{app:coordtrafo}) and on the $D$~dimensional harmonic
expansion of axisymmetric tensors (Appendix \ref{appintegrals}).

\section{Gravitational wave extraction in $D$ dimensional axially symmetric space-times\label{gwaves}}
\subsection{Coordinate frames}
In the approach developed in Paper I, we perform a dimensional reduction by isometry on the $(D-4)$-sphere $S^{D-4}$, in
such a way that the $D$~dimensional vacuum Einstein equations are rewritten as 
an effective $3+1$ dimensional time evolution problem with source terms that involve a scalar field. The
evolution equations are expressed in the Baumgarte-Shapiro-Shibata-Nakamura
(BSSN) formulation \cite{Shibata:1995we,Baumgarte:1998te}, and numerically implemented using a modification of the
\textsc{Lean} code \cite{Sperhake:2006cy}.

In Paper I we considered different generalizations of ``axial symmetries'' to higher dimensions: 
either $D\ge5$ dimensional space-times with $SO(D-2)$ isometry group, 
or $D\ge6$ dimensional space-times with $SO(D-3)$ isometry group. In this
work we only study the former case, which allows us to model head-on collisions of nonspinning BHs; 
we dub hereafter these space-times as \textit{axially symmetric}. 
Although the corresponding symmetry manifold is the $(D-3)$-sphere $S^{D-3}$, 
the quotient manifold in our dimensional reduction is its submanifold $S^{D-4}$. 
The coordinate frame in which the numerical simulations are performed is
\begin{equation}
(x^\mu,\phi^1,\dots,\phi^{D-4})=
(t,x,y,z,\phi^1,\dots,\phi^{D-4}) \, ,
\label{frame1}
\end{equation}
where the angles $\phi^1,\dots,\phi^{D-4}$ describe the quotient manifold $S^{D-4}$ 
and do not appear explicitly in the simulations.
Here, $z$ is the symmetry axis, i.e.\ the collision line.

In the frame \eqref{frame1}, the space-time metric has the form (cf. Eqs.~(2.14) and (2.21) of
Paper I)
\begin{align}
ds^2=&g_{\mu\nu}(x^\alpha)dx^\mu dx^\nu+\lambda(x^\mu)d\Omega_{D-4} \nonumber\\
    =&-\alpha^2dt^2+\gamma_{ij}(dx^i+\beta^idt)(dx^j+\beta^jdt)     \nonumber\\
     &+\lambda(x^\mu)d\Omega_{D-4} \, ,\label{totalmetric}
\end{align}
where $x^\mu=(t,x^i)$, $\lambda(x^\mu)$ is a scalar field and $\alpha,\beta^i$ are the lapse function and the shift
vector, respectively. It is worth noting that, although in $D=4$ a general axially symmetric space-time has
nonvanishing mixed components of the metric (like $g_{t\phi}$), in $D\ge5$ such components vanish in an appropriate
coordinate frame, as we have shown in Paper I.

With an appropriate transformation of the four dimensional coordinates $x^\mu$, the residual symmetry left after the
dimensional reduction on $S^{D-4}$ can be made manifest: $x^\mu\rightarrow (x^{\bar\mu},\theta)$ ($\bar\mu=0,1,2$),
\begin{equation}
g_{\mu\nu}(x^\alpha)dx^\mu dx^\nu=g_{\bar\mu\bar\nu}(x^{\bar\alpha})dx^{\bar\mu}dx^{\bar\nu}+g_{\theta\theta}(x^{\bar\alpha})d\theta^2
\label{thetametric}
\end{equation}
and
\begin{equation}
\lambda(x^\mu)=\sin^2\theta g_{\theta\theta}(x^{\bar\alpha})\,,\label{lambdatheta}
\end{equation}
so that Eq.~\eqref{totalmetric} takes the form $ds^2=g_{\bar\mu\bar\nu}dx^{\bar\mu}dx^{\bar\nu}+g_{\theta\theta}d\Omega_{D-3}$,
as discussed in Paper I.

To extract the gravitational waves far away from the symmetry axis
we employ the KI formalism
\cite{Kodama:2003jz}, which generalizes the Regge-Wheeler-Zerilli \cite{Regge:1957td,Zerilli:1970se} approach to higher dimensions. 
We require that the space-time, far away from the BHs, is approximately
spherically symmetric. Note, that spherical symmetry in $D$~dimensions means symmetry with respect to rotations on
$S^{D-2}$; this is an approximate symmetry which holds asymptotically
far away from the axis
and which is manifest in the coordinate frame:
\begin{equation}
(x^{a},\bar\theta,\theta,\phi^1,\dots,\phi^{D-4})=(t,r,\bar\theta,\theta,\phi^1,\dots,\phi^{D-4}) \,.
\label{frame2}
\end{equation}
Note that $x^{a}=t,r$ and that we have introduced polar-like coordinates $\bar\theta,\theta\in[0,\pi]$ 
to ``build up'' the manifold $S^{D-2}$ in the background, together with a radial spherical coordinate $r$, which
  is the areal coordinate in the background.

The coordinate frame \eqref{frame2} is defined in such a way that the metric 
can be expressed as a stationary background $(ds^{(0)})^2$ (i.e., the Tangherlini metric) 
plus a perturbation $(ds^{(1)})^2$ which decays faster than $1/r^{D-3}$ for large $r$:
\begin{align}
(ds^{(0)})^2=&g^{(0)}_{ab}dx^{a}dx^{b}+r^2d\Omega_{D-2} \nonumber\\
            =&g_{tt}^{(0)}dt^2+g_{rr}^{(0)}dr^2+r^2d\Omega_{D-2}\nonumber\\
            =&g_{tt}^{(0)}dt^2+g_{rr}^{(0)}dr^2+r^2\left(d\bar\theta^2+\sin^2\bar\theta 
d\Omega_{D-3}\right)
\nonumber\\
            =&-\left(1-\frac{r_S^{D-3}}{r^{D-3}}\right)dt^2
               +\left(1-\frac{r_S^{D-3}}{r^{D-3}}\right)^{-1}dr^2 \nonumber\\
              &+r^2\left[d\bar\theta^2+\sin^2\bar\theta
\left(d\theta^2+\sin^2\theta d\Omega_{D-4}\right)\right] \, ,\label{ds0}\\
(ds^{(1)})^2=& h_{ab}dx^{a}dx^{b}+h_{a\bar\theta}dx^{a}d\bar\theta
              +h_{\bar\theta\bar\theta}d\bar\theta^2 \nonumber\\
             &+h_{\theta\theta}d\Omega_{D-3}\,.\label{ds1}
\end{align}
Here, the Schwarzschild radius $r_S$ replaces the parameter $\mu$ used in Paper I and is 
related to the Arnowitt-Deser-Misner mass $M$
by
\begin{equation}
r_S^{D-3} = \frac{16\pi M}{(D-2){\cal A}_{D-2}} \, ,\label{defadm}
\end{equation}
where ${\cal A}_{D-2}$ is the area of the $(D-2)$-sphere (see Eq.~\eqref{areasphere}). For instance, $r_S=2M$ in $D=4$
and $r_S=\sqrt{8M/(3\pi)}$ in $D=5$. 

When we define the coordinate frame~\eqref{frame2}, we also require that the coordinate $\theta$ in this frame 
coincides with the coordinate  $\theta$ appearing in Eq.~\eqref{thetametric}. With this choice, the axial symmetry of the space-time implies that
\begin{equation}
h_{a\theta}=h_{\bar\theta\theta}=0\,,
\label{axialsymh}
\end{equation}
as in Eq.~\eqref{ds1}, and $\lambda=\sin^2\theta g_{\theta\theta}$, i.e. Eq.~\eqref{lambdatheta}.

The transformation from the coordinates $x^\mu=(t,x,y,z)$ in which the numerical simulation is implemented
to the coordinates $(x^{a},\bar\theta,\theta)=(t,r,\bar\theta,\theta)$
in which the wave extraction is performed is given by
\begin{align} 
x&=R\sin\bar\theta\cos\theta \, , \\ 
y&=R\sin\bar\theta\sin\theta \, ,\\ 
z&=R\cos\bar\theta \, ,
\label{transfc}
\end{align}
where $R = \sqrt{x^2 + y^2 + z^2}$
and by the reparametrization of the radial coordinate
\begin{equation}
R=R(r)\label{transfr}\,.
\end{equation}

We note that Eqs.~\eqref{transfc}, \eqref{transfr} correctly transform the three-metric $\gamma_{ij}$ describing 
the initial data
\begin{equation}
\gamma_{ij}dx^idx^j=\psi^{\frac{4}{D-3}}(dR^2+R^2(d{\bar\theta}^2+\sin^2\bar\theta d\theta^2))\,,\label{defgij2}
\end{equation}
where, as discussed in Paper I, we choose Brill-Lindquist initial data,
\begin{equation}
\psi = 1 + \frac{r_{S,1}^{D-3}}{4 r_1^{D-3}} +  \frac{r_{S,2}^{D-3}}{4 r_2^{D-3}}\,,
\label{BLinitdata}
\end{equation}
in order to simulate a head-on collision of BHs starting from rest.
Here, $r_{S,i}$ and $r_i$ denote the Schwarzschild radius and 
the initial position of the $i$-th BH, respectively.
Far away from the axis the conformal factor is given by
$\psi\rightarrow1+\hbox{ perturbations}$. Therefore, the splitting
of the metric into a Tangherlini background plus a perturbation, Eqs.~\eqref{ds0}-\eqref{ds1},
can be recovered on the initial time-slice, if we define 
the reparametrization~\eqref{transfr} appropriately.

Our guess is that the transformation~\eqref{transfc}, \eqref{transfr} yields the
``Tangherlini+perturbation'' splitting~\eqref{ds0}, \eqref{ds1} during the entire evolution of the system.
This statement can be checked
numerically by verifying the following relations (see Appendix~\ref{appintegrals}):
\begin{align}
\label{eq:testgtt}
\mathcal{G}_{tt} & \equiv \frac{1}{K^{0D}\pi} \int_0^{\pi} d\bar{\theta} \sin^{D-3}\bar{\theta}
  \int_0^{\pi} d \theta  g_{tt}(\bar{\theta},\theta) - g_{tt}^{(0)} = 0\,, \\
\label{eq:testgtR}
\mathcal{G}_{tr} & \equiv  \frac{1}{K^{0D}\pi} \int_0^{\pi} d\bar{\theta} \sin^{D-3}\bar{\theta} 
  \int_0^{\pi} d \theta  g_{tR}(\bar{\theta},\theta)  =0\,, \\
\label{eq:testgRR}
\mathcal{G}_{rr} & \equiv  
\frac{1}{K^{0D}\pi} \int_0^{\pi} d\bar{\theta} \sin^{D-3}\bar{\theta} 
  \int_0^{\pi} d \theta  g_{RR}(\bar{\theta},\theta) - g_{rr}^{(0)} = 0\,,
\end{align} 
where $K^{0D}=\int_0^\pi d\bar\theta(\sin\bar\theta)^{D-3}$, together with the axisymmetry
conditions~\eqref{lambdatheta}, \eqref{axialsymh}.
As we will discuss in Section~\ref{numres}, Eqs.~\eqref{lambdatheta}, \eqref{axialsymh}, \eqref{eq:testgtt}--\eqref{eq:testgRR} are indeed satisfied with high accuracy throughout the numerical evolution.
The preservation of the above identities during the numerical evolution  justifies also
the identification of the time coordinate in the numerical and wave extraction frames, 
and our use of the KI formalism. 

Finally, Eqs.~(\ref{totalmetric}), (\ref{ds0}), (\ref{ds1}) yield the $3+1$ splitting
\begin{align}
ds^2=& (ds^{(0)})^2 + (ds^{(1)})^2 \nonumber\\
    =& g_{\bar\mu\bar\nu}dx^{\bar\mu}dx^{\bar\nu}+(r^2\sin^2 \bar \theta
      +h_{\theta\theta})d\Omega_{D-3} \nonumber\\
    =& g_{\bar\mu\bar\nu}dx^{\bar\mu}dx^{\bar\nu}+(r^2\sin^2 \bar \theta+h_{\theta\theta}) \nonumber\\
     & \times(d\theta^2+\sin^2\theta d\Omega_{D-4})\nonumber\\
\label{eq:lineelement}
    =&-\alpha^2 dt^2+\gamma_{ij}(dx^i+\beta^idt) \nonumber\\
     & \times(dx^j+\beta^jdt)+\lambda d\Omega_{D-4} \, ,
\end{align}
where $x^{\bar\mu}=(t,r,\bar\theta)$.
With the $3+1$ splitting, the axisymmetry conditions~\eqref{lambdatheta},
\eqref{axialsymh} take the form
\begin{equation}
\lambda=\gamma_{\theta\theta}\sin^2\theta\,,~~~~~\gamma_{R\theta}=\gamma_{\bar\theta\theta}
=\beta^\theta=0\,.\label{checkax}
\end{equation}
The variable $r$ can be determined from the angular components of the metric (\ref{eq:lineelement}), by averaging out
$h_{\bar\theta\bar\theta}$, $h_{\theta\theta}$ (see Appendix \ref{appintegrals});
its explicit expression is given by
\begin{align}
\label{eq:arealradius}
(r(R))^2=& \frac{1}{(D-2)K^{0D}}\int_0^\pi d\bar\theta
           \left[\gamma_{\bar\theta\bar\theta}(\sin\bar\theta)^{D-3} \right. \nonumber\\
         & \left. +(D-3)\gamma_{\theta\theta}(\sin\bar\theta)^{D-5}\right]\,.
\end{align}
As we will discuss in Section~\ref{numres}, we find that the areal radius $r$ is very close to $R$.  
\subsection{Harmonic expansion}
In the KI formalism \cite{Kodama:2003jz} (see also \cite{Kodama:2000fa}), the  background  space-time has the form
(\ref{ds0})
\begin{align}
(ds^{(0)})^2 =& g^{(0)}_{AB}dx^Adx^B =g^{(0)}_{ab}dx^{a}dx^{b}+r^2d\Omega_{D-2} \nonumber\\
             =& g^{(0)}_{ab}dx^{a}dx^{b}+r^2\gamma_{\bar i\bar j}d\phi^{\bar i}d\phi^{\bar j} \, ,
\end{align}
i.e. the Tangherlini metric, where the $x^A$ coordinates refer to the full space-time. 
The space-time perturbations can be decomposed into spherical harmonics on
the $(D-2)$-sphere $S^{D-2}$. They are functions of the $D-2$ angles $\phi^{\bar i}=(\bar\theta,\theta,\phi^1,\dots,\phi^{D-4})$.  We denote  
the metric of $S^{D-2}$ by $\gamma_{\bar i\bar j}$, and with a subscript $_{:\bar i}$ the covariant derivative with
respect to this metric. Finally, we denote the covariant derivative with respect to the
metric $g^{(0)}_{ab}$ with a subscript $_{|a}$.

As discussed in \cite{Kodama:2003jz}, there are three types of spherical harmonics:
\begin{itemize}
\item The scalar harmonics ${\cal S}(\phi^{\bar i})$, which are
solutions of
\begin{equation}
\Box{\cal S}=\gamma^{\bar i\bar j}{\cal S}_{:\bar i\bar j}=-k^2{\cal S}\label{eqscal} \, ,
\end{equation}
with $k^2=l(l+D-3)$, $l=0,1,2,\dots$ . 
The scalar harmonics ${\cal S}$ depend on the integer $l$ and on other
indices; we leave such dependence implicit. We also define
\begin{align}
{\cal S}_{\bar i}=&-\frac{1}{k}{\cal S}_{,\bar i}\,,\,\,
{\cal S}_{\bar i\bar j}=\frac{1}{k^2}{\cal S}_{:\bar i\bar j}+\frac{1}{D-2}\gamma_{\bar i\bar j}{\cal S} \, .
\label{defharm}
\end{align}
Observe that $\gamma^{\bar i\bar j}{\cal S}_{\bar i\bar j}=0$.

Each harmonic mode of the metric perturbation $\delta g_{MN}=h_{MN}$ can be decomposed as
\begin{eqnarray}
\delta g_{ab}=h_{ab}
&=&f_{ab}{\cal S}\label{mpert0} \, , \\
\delta g_{a\bar i}=h_{a\bar i}
&=&rf_{a}{\cal S}_{\bar i}\label{mpert1}\, ,\\
\delta g_{\bar i\bar j}=h_{\bar i\bar j}
&=&2r^2(H_L\gamma_{\bar i\bar j}{\cal S}+H_T{\cal S}_{\bar i\bar j})\label{mpert2} \, ,
\end{eqnarray}   
where $f_{ab}$, $f_{a}$, $H_L$, $H_T$ are functions of $x^{a}=(t,r)$.
Note, that in each of these expressions there is a sum over the indices of the harmonic. 

For $l>1$, the metric perturbations can be expressed in terms of the following gauge-invariant variables
\cite{Kodama:2000fa}
\begin{eqnarray}
\label{eq:func1}
F&=&H_L+\frac{1}{D-2}H_T+\frac{1}{r}X_{a} r^{|a}\, ,\nonumber\\
\label{eq:func2}
F_{ab}&=&f_{ab}+X_{a|b}+X_{b|a} \, ,\label{gi1}
\end{eqnarray}
where we have defined
\begin{equation}
X_{a}=\frac{r}{k}\left(f_{a}+\frac{r}{k}H_{T|a}\right)\,.\label{gi2}
\end{equation}
\item The vector harmonics ${\cal V}_{\bar i}(\phi^{\bar i})$, solutions of 
\begin{equation}
\gamma^{\bar i\bar j}{\cal V}_{\bar k:\bar i\bar j}=-k_V^2{\cal V}_{\bar k} \, ,
\end{equation} 
with $k_V^2=l(l+D-3)-1$, $l=1,2,\dots$. These harmonics satisfy the relation
\begin{equation}
{\cal V}^{\bar i}_{:\bar i}=0\,.\label{divV}
\end{equation}

The harmonic expansion of the corresponding metric perturbations is given by Eqs.~(\ref{mpert1})-(\ref{mpert2}),
with ${\cal S}_{\bar i}$ replaced by ${\cal V}_{\bar i}$, ${\cal S}_{\bar i\bar j}$ replaced by
\begin{equation}
{\cal V}_{\bar i\bar j}=-\frac{1}{2k_V}(V_{\bar i:\bar j}+V_{\bar j:\bar i})\,,
\end{equation}
and $H_L=0$.
\item The tensor harmonics ${\cal T}_{\bar i\bar j}(\phi^{\bar i})$, solutions of 
\begin{equation}
\gamma^{\bar i\bar j}{\cal T}_{\bar r\bar s:\bar i\bar j}=-k_T^2{\cal T}_{\bar r\bar s} \, ,
\end{equation} 
with $k_T^2=l(l+D-3)-2$, $l=1,2,\dots$. These harmonics satisfy,
\begin{equation}
\gamma^{\bar i\bar j}{\cal T}_{\bar i\bar j}=0\,,
~~~~~{\cal T}^{:\bar i\bar j}_{:\bar j}=0\,.\label{divT}
\end{equation}
In the $D=4$ case they vanish. The harmonic expansion of the corresponding metric perturbations is given by
(\ref{mpert2}), with ${\cal S}_{\bar i\bar j}$ replaced by ${\cal T}_{\bar i\bar j}$ and $H_L=0$.
\end{itemize}
\subsection{Implementation of axisymmetry}\label{implax}
In an axially symmetric space-time, the metric perturbations are symmetric with respect to $S^{D-3}$. Therefore, the
harmonics in the expansion of $h_{MN}$ depend only on the angle $\bar\theta$ (which does not belong to
$S^{D-3}$). Furthermore, since there are no off-diagonal terms in the metric (cf. Paper I), the only nonvanishing
$g_{a\bar i}$ components are $g_{a\bar\theta}$; the only components $g_{\bar i\bar j}$ are either
proportional to $\gamma_{\bar i\bar j}$, or all vanishing but $g_{\bar\theta\bar\theta}$. This implies that only
scalar spherical harmonics can appear in the expansion of the metric perturbations. Indeed, if
\begin{equation}
{\cal V}^{\bar i}=({\cal V}^{\bar\theta},0,\dots,0)\,,~~~~~{\cal V}^{\bar i}={\cal V}^{\bar i}(\bar\theta) \, ,
\end{equation}
then Eq.~(\ref{divV}) gives
\begin{equation}
{\cal V}^{\bar i}_{:\bar i}={\cal V}^{\bar\theta}_{,\bar\theta}=0~~~\Rightarrow~~~
{\cal V}^{\bar\theta}=0~~\Rightarrow~~~{\cal V}^{\bar i}=0\,.
\end{equation}
Similarly, from Eq.~(\ref{divT}) we obtain ${\cal T}_{\bar i\bar j}=0$.

The scalar harmonics, solutions of Eq.~(\ref{eqscal}) and which depend only on the coordinate $\bar\theta$, are given by the
Gegenbauer polynomials $C_l^{(D-3)/2}$, as discussed in Refs.~\cite{Berti:2003si,Cardoso:2002ay,Yoshino:2005ps}; writing explicitly the index $l$, they take the form
\begin{equation}
{\cal S}_l(\bar\theta)=(K^{lD})^{-1/2}C_l^{(D-3)/2}(\cos\bar\theta) \, , \label{gp}
\end{equation} 
where the normalization $K^{lD}$ is chosen such that
\begin{equation}
\int d\Omega^{D-2}{\cal S}_l{\cal S}_{l'}=\delta_{ll'}\,,~~~
\int d\Omega^{D-2}{\cal S}_{l\,,\bar\theta}{\cal S}_{l'\,,\bar\theta}=\delta_{ll'}k^2 \ ,\label{normalK}
\end{equation}
and $k^2=l(l+D-3)$ (see Appendix \ref{appintegrals}). By computing ${\cal S}_{l\,\bar i}$, ${\cal S}_{l\,\bar i\bar j}$ from
Eqs.~\eqref{defharm} (using Eq.~\eqref{eqscal}) we find
\begin{align}
{\cal S}_{l\,\bar\theta\bar\theta}&=\frac{D-3}{k^2(D-2)} {\cal W}_l \, , \\
{\cal S}_{l\,\theta\theta}&=-\frac{\sin^2\bar\theta}{k^2(D-2)}{\cal W}_l\,,
\end{align}
where we have defined
\begin{equation}
{\cal W}_l(\bar\theta) ={\cal S}_{l\,,\bar\theta\bar\theta}-\cot\bar\theta{\cal S}_{l\,,\bar\theta}\,.
\end{equation}
Therefore, the metric perturbations are given by
\begin{align}
h_{ab} =& f^l_{ab}{\cal S}_l(\bar\theta)\, , \\
h_{a\bar\theta} =& rf^l_{a}{\cal S}_l(\bar\theta)_{\bar\theta}=
-\frac{1}{k} rf^l_{a}{\cal S}_l(\bar\theta)_{,\bar\theta}\,  ,\\
h_{\bar\theta\bar\theta} =& 2r^2(H^l_L {\cal S}_l(\bar\theta)
                           +H^l_T {\cal S}_l(\bar\theta)_{\bar\theta\bar\theta}) \nonumber\\
                         =& 2r^2\left(H^l_L {\cal S}_l(\bar\theta)
                           +H^l_T \frac{D-3}{k^2(D-2)}{\cal W}_l(\bar\theta)\right)\, , \\
h_{\theta\theta} =& 2r^2(H^l_L\sin^2\bar\theta {\cal S}_l(\bar\theta)
                   +H^l_T {\cal S}_l(\bar\theta)_{\theta\theta}) \nonumber\\
                 =& 2r^2\sin^2\bar\theta\left(H^l_L {\cal S}_l(\bar\theta)
                   -H^l_T \frac{1}{k^2(D-2)}{\cal W}_l(\bar\theta)\right) \, .
\end{align}\label{mpert}
The quantities $f_{ab}$, $f_{a}$, $H_L$, $H_T$ are (see Appendix \ref{appintegrals}):
\begin{align}
f^l_{ab}(t,r) =& \frac{{\cal A}_{D-3}}{\sqrt{K^{lD}}}\int_0^{\pi}
                 d\bar\theta (\sin\bar\theta)^{D-3}h_{ab}  C_l^{(D-3)/2}\,  ,\\
f_{a}(t,r) =&-\frac{1}{\sqrt{l(l+D-3)}r}\frac{{\cal A}_{D-3}}{\sqrt{K^{lD}}} \nonumber\\
            & \times\int_0^{\pi}d\bar\theta (\sin\bar\theta)^{D-3}
                    h_{a \bar \theta }C_{l\,,\bar\theta}^{(D-3)/2}(\cos{\bar\theta})\, , \\
H_L(t,r) =& \frac{1}{2(D-2)r^2}\frac{{\cal A}_{D-3}}{\sqrt{K^{lD}}}\int_0^{\pi}
            d\bar\theta (\sin\bar\theta)^{D-3} \nonumber\\
          & \times \left[h_{\bar\theta\bar\theta}+\frac{D-3}{\sin^2\bar\theta}h_{\theta\theta}\right] 
            C_l^{(D-3)/2}(\cos{\bar\theta})\,  , \\
H_T(t,r) =& \frac{1}{2r^2(k^2-D+2)}\frac{{\cal A}_{D-3}}{\sqrt{K^{lD}}}
            \int_0^{\pi}d\bar\theta (\sin\bar\theta)^{D-3} \nonumber\\
          & \times \left[h_{\bar\theta\bar\theta}-\frac{1}{\sin^2\bar\theta}h_{\theta\theta}\right]W_l(\bar\theta)\,,
\end{align}\label{metricrelation}
where $h_{ab} = h_{ab}(t,r,\bar\theta)$, $h_{a\bar\theta} = h_{a\bar\theta}(t,r,\bar\theta)$,
$h_{\bar\theta \bar\theta} = h_{\bar\theta \bar\theta}(t,r,\bar\theta)$, 
$h_{\theta\theta} = h_{\theta \theta}(t,r,\bar\theta)$ and 
$C_l^{(D-3)/2} = C_l^{(D-3)/2}(\cos{\bar\theta})$.

In terms of these quantities, using Eqs.~(\ref{gi1}), (\ref{gi2}), we get the gauge-invariant quantities $F$, $F_{  ab}$.

As we have discussed above, this approach has been developed for $D>4$, since in $D=4$ the off-diagonal terms
$g_{t\phi}$, $g_{r\phi}$ are not vanishing in general axially symmetric space-times. However, we can extend our
framework to $D=4$ if we restrict ourselves to axially symmetric space-times with $g_{t\phi}=g_{r\phi}=0$. In this way,
we can test our formalism by comparing our results to the existing literature. For instance, we note that in $D=4$ the
perturbation functions are related to the expressions in Ref.~\cite{Sperhake:2005uf}, with the identifications
\begin{eqnarray}
\label{eq:compPF1}
f^l_{ab}&=&H_0,\,H_1,\,H_2\,,\\
\label{eq:compPF2}
-\frac{r}{k}f^l_{a}&=&h_0,\,h_1\,,\\
\label{eq:compPF3}
\frac{2H_T}{k^2}&=&G\,,\\
\label{eq:compPF4}
2H_L+H_T&=&K\,.
\end{eqnarray}
We also remark that in the transverse-traceless gauge, only $H_T$ is nonvanishing, but in a generic gauge (like the one
used in the numerical simulations) all these quantities are in principle nonvanishing.
\subsection{Extracting gravitational waves at infinity}\label{master}
In the KI framework, the emitted gravitational waves are described by the master function $\Phi$. To compute $\Phi$ in
terms of the gauge-invariant quantities $F$, $F_{ab}$ one should perform a Fourier transform or a time
integration (see \cite{Kodama:2003jz}). This can be avoided if we compute directly $\Phi_{,t}$, given by\footnote{Note
  that there is a factor $r$ missing in Eq.~(3.15) of Ref.~\cite{Kodama:2003jz}.}
\begin{equation} 
\label{eq:KIwavefunction}
\Phi_{,t}=(D-2)r^{(D-4)/2}\frac{-F^r_{~t}+2 r F_{,t}}{k^2-D+2+\frac{(D-2)(D-1)}{2}\frac{r_S^{D-3}}{r^{D-3}}} \, ,
\end{equation}
where $k^2=l(l+D-3)$. In the TT-gauge, the gravitational perturbation is described by $H_T$, which decays as
$r^{(D-2)/2}$ with increasing $r$, whereas the other perturbation functions have a faster decay (see
\cite{Berti:2003si}). In this gauge, the asymptotic behaviour of the master function is
\begin{equation}
\Phi\simeq\frac{2r^{(D-2)/2}H_T}{k^2}\,  ,\label{propfinite}
\end{equation}
and tends to an oscillating function with constant amplitude as $r\rightarrow\infty$.  The asymptotic behaviour of
$\Phi$ has been checked numerically (cf. Section \ref{numres}).

Writing the index $l$ explicitly, the energy flux in each $l-$multipole is \cite{Berti:2003si}
\begin{equation}
\label{eq:energyflux}
\frac{dE_l}{dt}=\frac{1}{32\pi}\frac{D-3}{D-2}k^2(k^2-D+2)(\Phi^l_{,t})^2\,.
\end{equation} 
The total energy emitted in the process is then
\begin{equation}
\label{eq:energyrad}
E=\sum_{l=2}^\infty\int_{-\infty}^{+\infty}dt\frac{dE_l}{dt}\,.
\end{equation}
%

\section{Head-on collision from rest in $D=4$}
\label{numres}
The numerical simulations of head-on collisions of equal-mass binaries
starting from rest have been performed with the \textsc{Lean}
code originally introduced in Ref.~\cite{Sperhake:2006cy}, modified along Sec.~III of Ref.~\cite{Sperhake:2007gu}
and adapted to higher dimensional space-times in Paper I.
The \textsc{Lean} code is based on the \texttt{Cactus}
computational toolkit \cite{cactus} and uses the \texttt{Carpet}
mesh refinement package \cite{Schnetter:2003rb, carpet},
the apparent horizon finder \texttt{AHFinderDirect}
\cite{Thornburg:1995cp,Thornburg:2003sf} and the puncture initial
data solver of Ref.~\cite{Ansorg:2004ds}.
Head-on collisions in four dimensional space-times have been
studied extensively in the literature and provide valuable
opportunities to calibrate the wave extraction formalism.
These tests are the subject of the remainder of this Section, while
we discuss our new results for five dimensional space-times
in Sec.~\ref{sec:5D} below.
\begin{table*}
\caption{\label{tab:setup}
  Grid structure and initial parameters of the head-on collisions
  starting from rest in $D=4$ and $D=5$.
  The grid setup is given in terms of the ``radii'' of the individual refinement levels, in units
  of $r_S$, as well as the resolution near
  the punctures $h$ (see Sec.~II E in \cite{Sperhake:2006cy} for details).
  $d$ is the initial coordinate  separation of the two punctures
  and $L$ denotes the proper initial separation.
}
\begin{tabular}{cccccc}
\hline
Run & $D$ & Grid Setup & $d/r_S$ & $L/r_S$ \\ \hline 
HD4$_c$ & 4 & $\{(128,64,32,16,8)\times(1,0.5,0.25),~h=r_S/80\}$ & $5.257$ & $7.154$ \\ 
HD4$_m$ & 4 & $\{(128,64,32,16,8)\times(1,0.5,0.25),~h=r_S/88\}$ & $5.257$ & $7.154$ \\ 
HD4$_f$ & 4 & $\{(128,64,32,16,8)\times(1,0.5,0.25),~h=r_S/96\}$ & $5.257$ & $7.154$ \\ 
HD5a & 5 & $\{(256,128,64,32,16,8,4)\times(0.5,0.25),~h=r_S/84\}$ & $1.57$ & $1.42$ & \\ 
HD5b & 5 & $\{(256,128,64,32,16,8,4)\times(0.5,0.25),~h=r_S/84\}$ & $1.99$ & $1.87$ \\ 
HD5c & 5 & $\{(256,128,64,32,16,8,4)\times(1,0.5),~h=r_S/84\}$ & $2.51$ & $2.41$ \\ 
HD5d & 5 & $\{(256,128,64,32,16,8,4)\times(1,0.5),~h=r_S/84\}$ & $3.17$ & $3.09$ \\ 
HD5e$_c$ & 5 & $\{(256,128,64,32,16,8)\times(2,1,0.5),~h=r_S/60\}$ & $6.37$ & $6.33$ \\ 
HD5e$_m$ & 5 & $\{(256,128,64,32,16,8)\times(2,1,0.5),~h=r_S/72\}$ & $6.37$ & $6.33$ \\ 
HD5e$_f$ & 5 & $\{(256,128,64,32,16,8)\times(2,1,0.5),~h=r_S/84\}$ & $6.37$ & $6.33$ \\ 
HD5f & 5 & $\{(256,128,64,32,16,8)\times(2,1,0.5),~h=r_S/84\}$ & $10.37$ & $10.35$ \\ 
\hline
\end{tabular}
\end{table*}

In order to test our implementation of the KI formalism
in $D=4$, we have simulated head-on collision
of an equal-mass, non spinning BH binary initially at rest.
The parameters used in this simulation are shown in
Table \ref{tab:setup}. 
This particular system is well understood
and enables us to compare our results derived from the KI formalism
against those obtained using both,
the Regge-Wheeler-Zerilli wave extraction and the Newman-Penrose
framework; cf. \cite{Sperhake:2005uf} and \cite{Sperhake:2006cy} 
for corresponding literature studies.

In order to perform these tests, we need to relate our master
function $\Phi$ of Sec.~\ref{master} to the variables used in
traditional four dimensional studies. Specifically, a straightforward
calculation shows that the Zerilli wavefunction $\bar\Phi$
adopted in Ref.~\cite{Sperhake:2005uf} for $l=2$ multipoles
and the outgoing Weyl scalar $\Psi_4$ used in \cite{Sperhake:2006cy} 
can be expressed in terms of $\Phi$ according to
\begin{eqnarray}
\bar\Phi &=&6\Phi\,,\label{rel1}\\
r \Psi_4&=&\sqrt{6}\Phi_{,tt}\label{rel2}\,.
\end{eqnarray}
Note that the imaginary part of $\Psi_4$ vanishes in the case of a head on collision, due to symmetry.

The resolution is $h=r_S/96$ for all results reported in this section except for
the convergence study in Sec.~\ref{raden} which also uses the lower resolutions
$h_c=r_S/80$ and $h_m=r_S/88$.\footnote{In order to ensure that our fundamental
  unit is of physical dimension length for all values of space-time dimension
  $D$, we believe it convenient to express our results in units of the radius
  $r_S$ (given by $r_S^{D-3} \equiv r^{D-3}_{S,1}+r^{D-3}_{S,2}$) of the
  ``total'' event horizon as opposed to the total BH mass $M$ commonly used in
  four dimensional numerical relativity. In $D=4$, of course, $r_S=2M$.}
Gravitational waves have been extracted at three different coordinate
  radii $R$ (cf. Eq.~\ref{transfr}), which we denote by $R_{\rm ex} =
  30\,r_S\,,40\,r_S\,,50\,r_S$.

\subsection{Tests on the numerical coordinates}
\label{sec:coordtestD4}
The procedure described in Section \ref{gwaves} assumes that
the numerical space-time consists of a small deviation from the
Schwarzschild-Tangherlini metric. In order to ensure that the
gravitational waves are extracted in
an appropriate coordinate system we perform a number of checks.
\begin{figure*}
\begin{center}
\begin{tabular}{cc}
\includegraphics[width=0.45\textwidth]{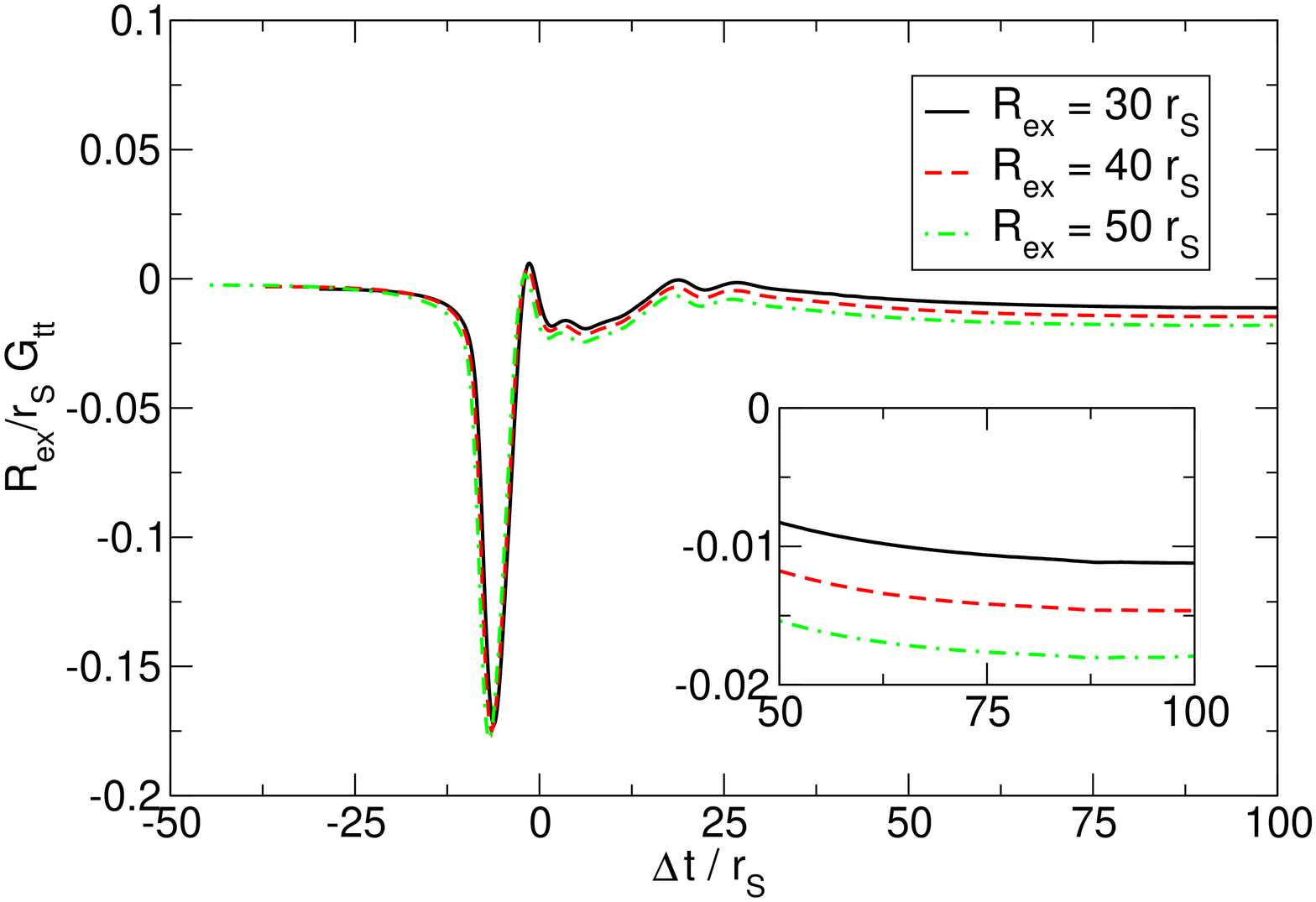} &
\includegraphics[width=0.45\textwidth]{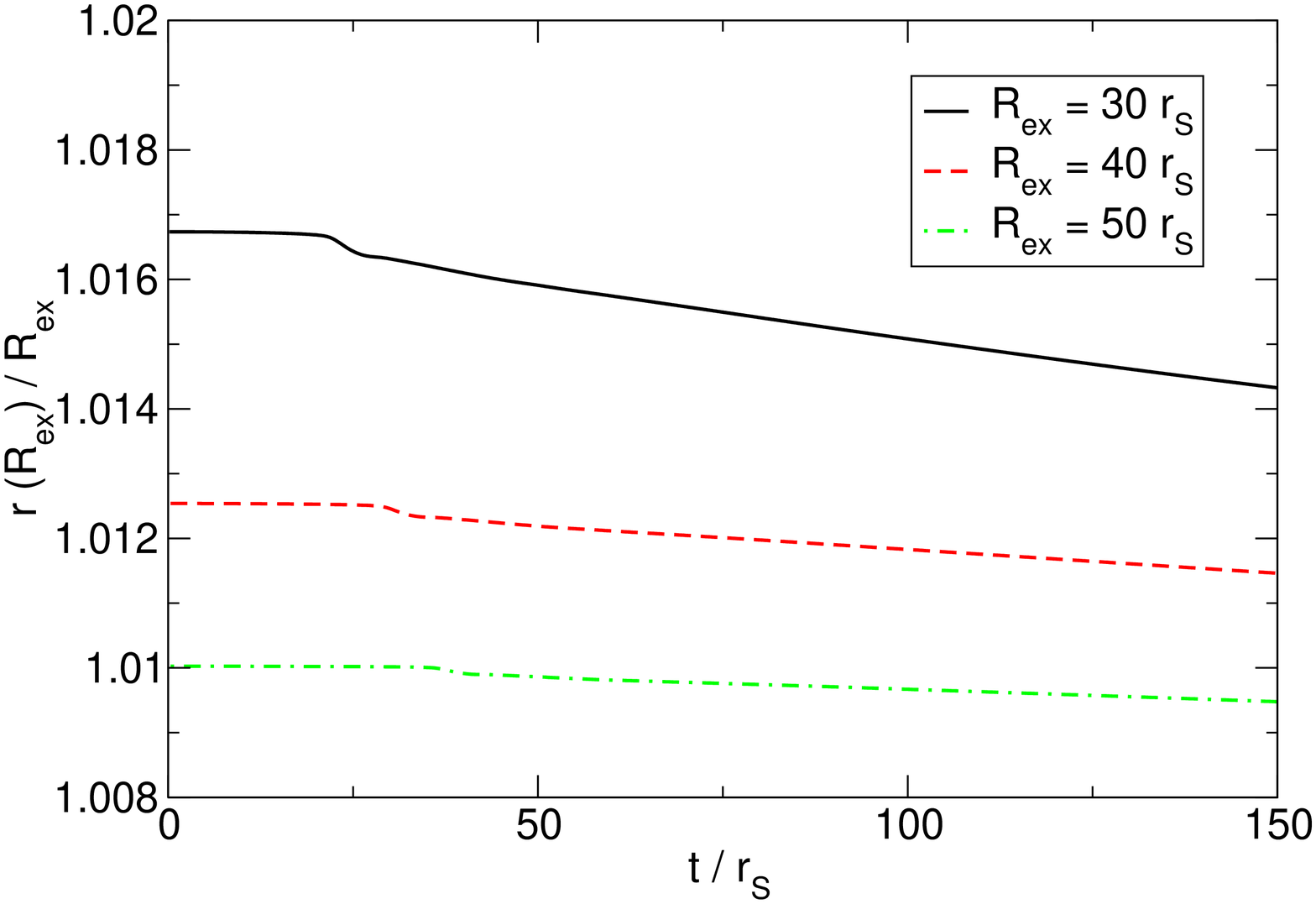}
\end{tabular}
\end{center}
\caption{\label{fig:coordinates} (Color online)
Left panel: $\mathcal{G}_{tt}$ calculated from Eq.~\eqref{eq:testgtt}
for $D=4$, at different extraction radii. This quantity has been shifted
in time to account for the different extraction radii and re-scaled
by the corresponding $R_{\rm ex}$. The late time behavior is shown in
the inset.
Right panel: time evolution of areal radius
(cf. (\ref{eq:arealradius}))  extracted at the radii $R_{\rm ex} = 30r_S$
(black solid line), $R_{\rm ex} = 40r_S$ (red dashed line) and $R_{\rm ex} =
50r_S$ (green dashed-dotted line).
}
\end{figure*}
We first test the relations \eqref{eq:testgtt}, \eqref{eq:testgtR}
and \eqref{eq:testgRR}. In Fig. \ref{fig:coordinates} we show
$\mathcal{G}_{tt}$, i.e., the difference between
the numerical $g_{tt}$, averaged over the extraction sphere
and the corresponding component of the assumed background metric.
Here we evaluate the background metric by assuming, as a first approximation,
that the Schwarzschild radius of the BH is $r_S=r_{S,1}+r_{S,2}$.

The deviation of the full 4-metric from the
Schwarzschild-Tangherlini background decreases as the extraction radius
increases. Indeed, a straightforward calculation shows that a deviation $\delta r_S$ of
the Schwarzschild radius from the background value leads to
$\mathcal{G}_{tt}\sim \delta r_S^{D-3}/r^{D-3}$,
i.~e.~$\delta r_S / r$ for $D=4$.
In the left panel of Fig.~\ref{fig:coordinates} 
we therefore show the deviation $\mathcal{G}_{tt}$ re-scaled by $r$. We further apply a time shift
to account for the different propagation time of the wave to reach the extraction radii. As shown
in the figure, the deviation from the Schwarzschild line element {\it is} small and decreases
$\sim 1/r$ in accordance with our expectation. We also note that a deviation $\delta r_S$ represents a monopole
perturbation of the background which decouples from the quadrupole
wave signal at perturbative order, so that its impact on our results
is further reduced.

In summary, we can give an uncertainty estimate for the approximation $r_S = r_{S,1} + r_{S,2}$ for the Schwarzschild radius of the final BH,
which ignores the energy loss through gravitational radiation. As demonstrated by the left panel of
Fig.~\ref{fig:coordinates}, at late times $|R_{\rm ex}/r_S\,{\cal G}_{tt}|\sim 0.01$, and, since
$r\simeq R_{\rm ex}$ (as we discuss below), we obtain the upper bound
\begin{equation}
\frac{\delta r_S}{r_S}\lesssim\frac{r}{r_S}{\cal G}_{tt}\sim 0.01\,.
\end{equation}
This crude analysis sets an upper bound of $\sim 1\%$ on the fraction of the centre of mass energy radiated as gravitational waves.
We further note that the close agreement between $g_{tt}$ and its Tangherlini counterpart 
implies that the time coordinate employed in the numerical simulation and the Tangherlini coordinate time coincide. 
By analysing $\mathcal{G}_{tr}$ and $\mathcal{G}_{rr}$ in the same
manner, we find that relations \eqref{eq:testgtt}-\eqref{eq:testgRR}
are satisfied with an accuracy of one part in $10^2$ throughout the evolution,
and one part in $10^3$ at late times, when the space-time consists of a single distorted black hole.

In practice, gravitational waves are extracted on spherical shells
of constant coordinate radius. The significance of the areal
radius associated with such a coordinate sphere in the context
of extrapolation of GW signals has been studied
in detail in Ref.~\cite{Boyle:2009vi}. For our purposes, the most
important question is to what extent gauge effects change the
areal radius (\ref{eq:arealradius}) of our extraction spheres.
For this purpose, we show its time evolution in the right panel of Fig. \ref{fig:coordinates}
for different values of $R_{\rm ex}$. The reassuring result is that the areal radius
exceeds its coordinate counterpart by about $1~\%$ at $R_{\rm ex}=50~r_S$
and remains nearly constant in time. 
\subsection{Waveforms}
%
\begin{figure*}
\begin{center}
\begin{tabular}{cc}
\includegraphics[width=0.45\textwidth]{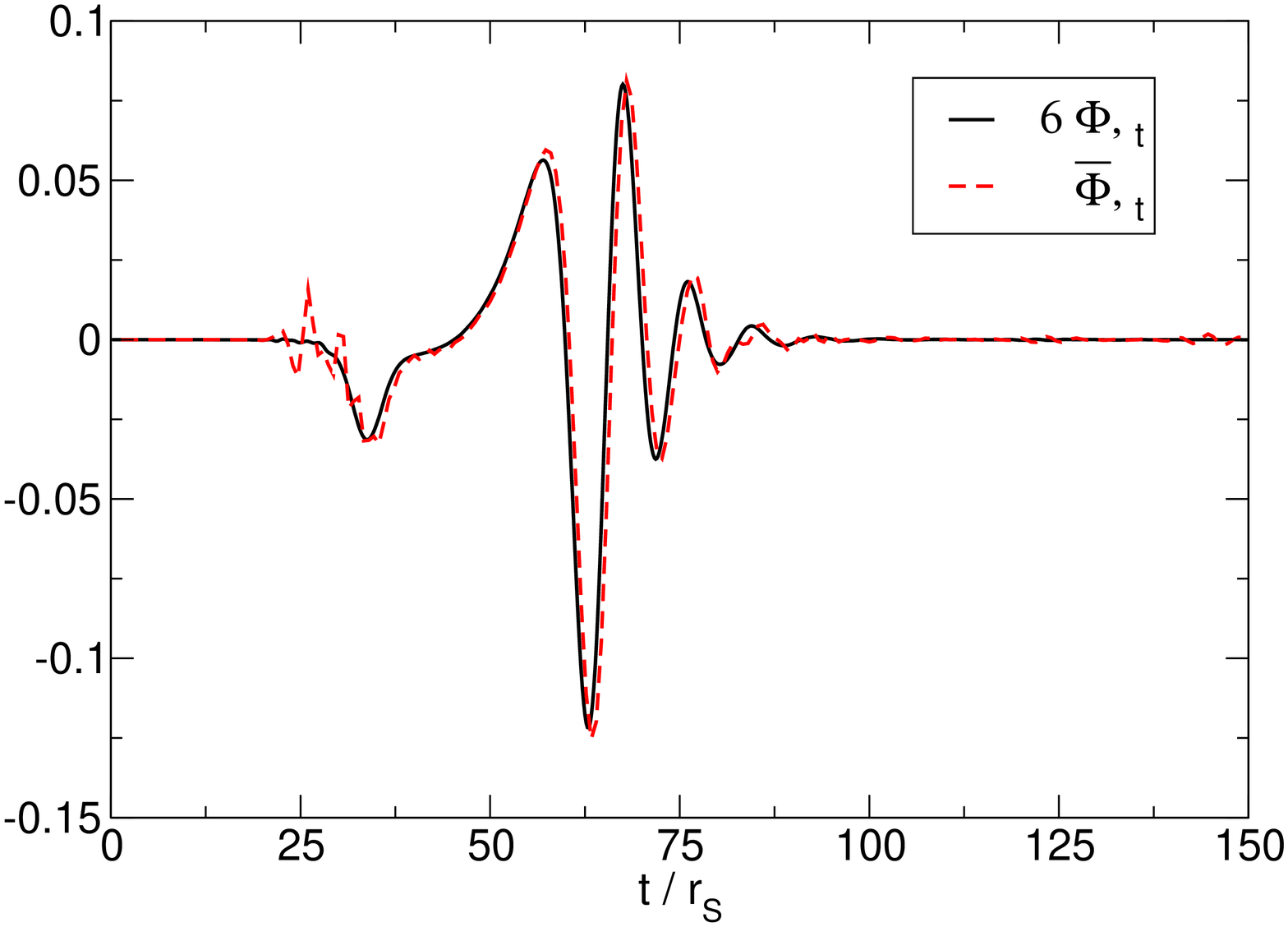} &
\includegraphics[width=0.45\textwidth]{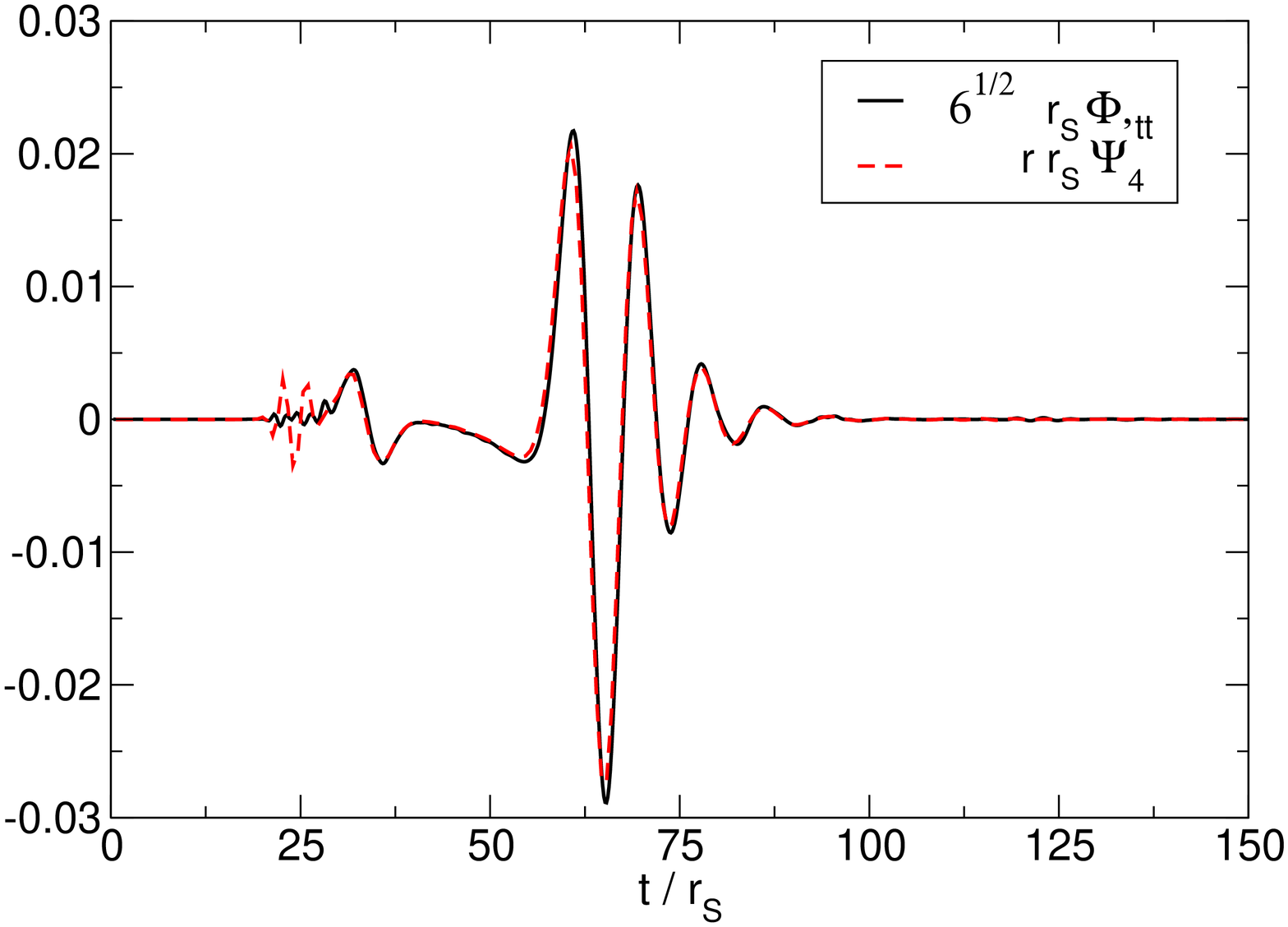} 
\end{tabular}
\end{center}
\caption{\label{fig:comparison1} (Color online) 
  Left panel: Time derivatives of the $l=2$ modes of the KI
  function $\Phi$ (black solid line), and of the Zerilli function
  $\bar\Phi$ (red dashed line) extracted for model HD4$_f$
  at $R_{\rm ex}=30r_S$.
  The KI function has been re-scaled by a constant factor (cf.~\eqref{rel1}) 
  which accounts for the different normalizations of both formulations. 
  Right panel: comparison of the second time derivative $\Phi_{,tt}$
  with the outgoing Newman-Penrose scalar $\Psi_4$ for the same model. 
  The KI wavefunction has been re-scaled 
  according to Eq.~\eqref{rel2}.
}
\end{figure*}
As a benchmark for our wave extraction, we compare our results
obtained with independent wave extraction tools; (i) the
explicitly four dimensional Zerilli formalism and (ii)
the Newman-Penrose scalars.
For this purpose we have evolved model HD$4_f$ and extracted the
Zerilli function according to the procedure described in
\cite{Sperhake:2005uf}
(see also Eqs.~(\ref{eq:compPF1}) - (\ref{eq:compPF4}) above)
and the Newman Penrose scalar $\Psi_4$ as summarized in
\cite{Sperhake:2006cy}. These are compared
with the KI wave function $\Phi_{,t}$ and its time derivative
$\Phi_{,tt}$ in Fig.~\ref{fig:comparison1}. Except for a small
amount of high frequency noise in the junk radiation at $t\approx 25 r_S$,
we observe excellent agreement between the different extraction methods.
\begin{figure}
\begin{center}
\begin{tabular}{cc}
\includegraphics[width=0.45\textwidth]{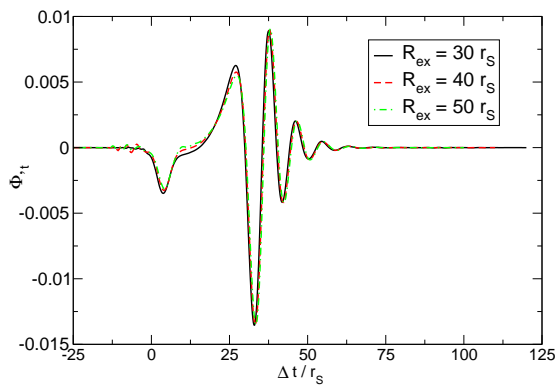}
\end{tabular}
\end{center}
\caption{\label{fig:Phir} (Color online) The $l=2$ component of the
  KI wave function $\Phi_{,t}$ extracted
  at the radii $R_{\rm ex} = 30r_S$ (black solid line), $R_{\rm ex} = 40r_S$
  (red dashed line) and $R_{\rm ex} = 50r_S$ (green dashed-dotted line). They
  have been shifted in time by the corresponding $R_{\rm ex}$.}
\end{figure}
Next we consider the dependence of the wave signal on the
extraction radius. 
In Fig.~\ref{fig:Phir} we show the $l=2$ component of $\Phi_{,t}$
extracted at three different radii and shifted in time by $R_{\rm ex}$.
As is apparent from the figure, the wave function shows little variation
with $R_{\rm ex}$ at large distances, in agreement with expectations.

A further test of the wave signal arises from its late-time behaviour which is dominated by the BH
ringdown \cite{Berti:2009kk}, an exponentially damped sinusoid of the 
form $e^{-{\rm i}\omega t}$, with $\omega$ being a characteristic frequency 
called quasinormal mode (QNM) frequency. Using well-known methods \cite{Berti:2005ys,Berti:2007dg,Berti:2009kk}, 
we estimate this frequency to be $r_S\,\omega\sim 0.746\pm0.002-{\rm i}\,(0.176\pm0.002)$.  
This can be compared with theoretical predictions from a linearized approach,
yielding $r_S\,\omega=0.747344-{\rm i}\,0.177925$.  
\begin{figure}
\begin{center}
\includegraphics[width=0.45\textwidth]{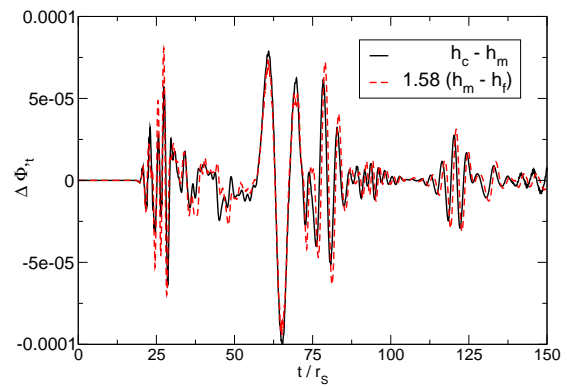}
\end{center}
\caption{\label{fig:convergence} (Color online) 
  Convergence analysis of the $l=2$ component of $\Phi_{,t}$ extracted at 
  $R_{\rm ex} = 30\,r_S$. We plot the differences between the low and medium resolution
  (black solid line) and medium and high resolution (red dashed line)
  run. The latter is re-scaled by the factor $Q=1.58$ expected for
  $4^{th}$ order convergence \cite{Witek:2010qc}. 
}
\end{figure}
Finally, we consider the numerical convergence of our results.
In Fig.~\ref{fig:convergence}, we plot the differences obtained
for $\Phi_{,t}$ extracted at $R_{\rm ex} = 30\,r_S$, using the different 
resolutions of the three models HD$4$ listed in Table~\ref{tab:setup}.
The differences thus obtained are consistent with $4^{th}$ order convergence. 
This implies a discretization error in the $l=2$ component of $\Phi_{,t}$ 
of about $4\%$ for the grid resolutions used in this work.

\subsection{Radiated energy}\label{raden}
\begin{figure*}
\begin{center}
\begin{tabular}{cc}
\includegraphics[width=0.45\textwidth]{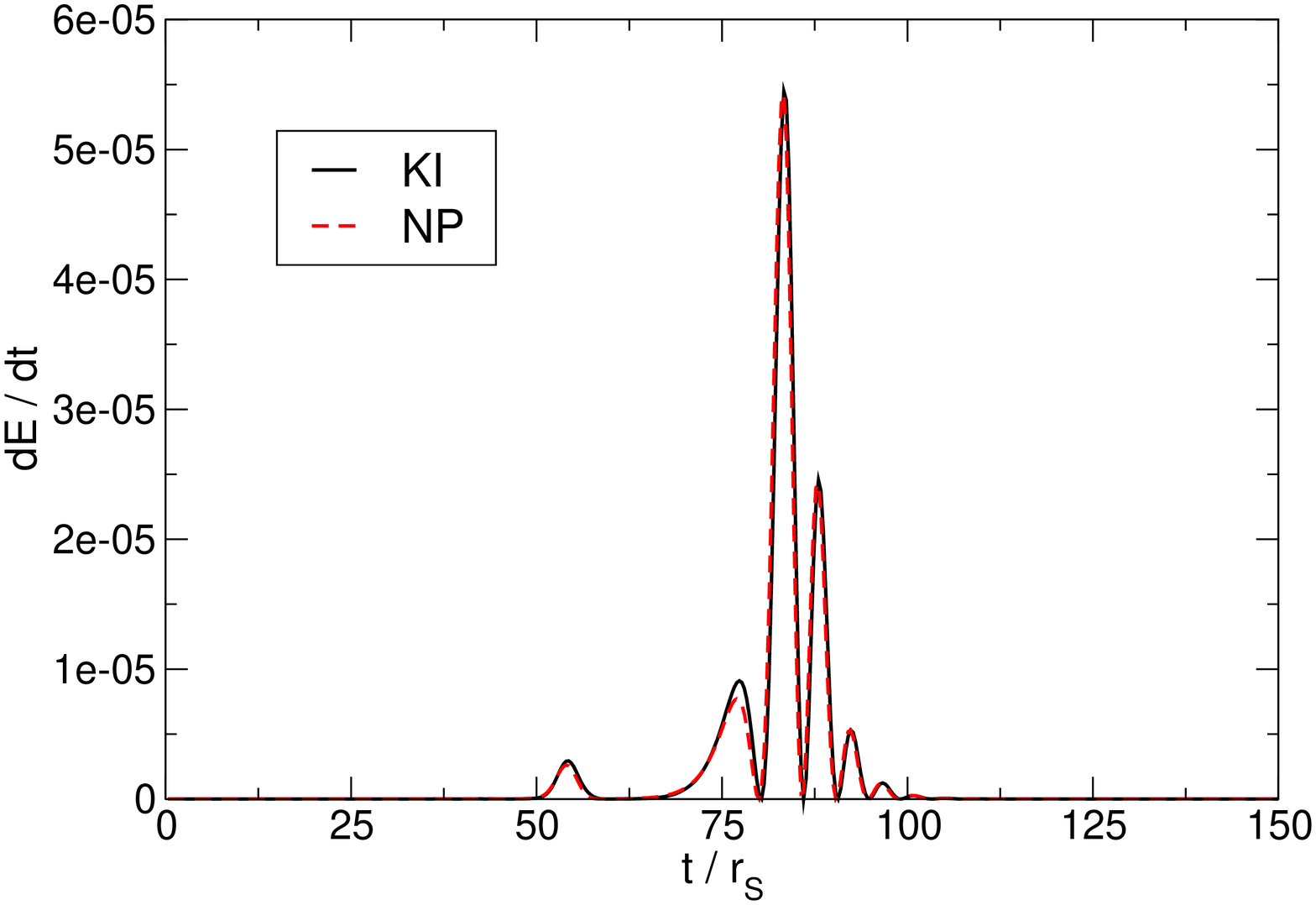} &
\includegraphics[width=0.45\textwidth]{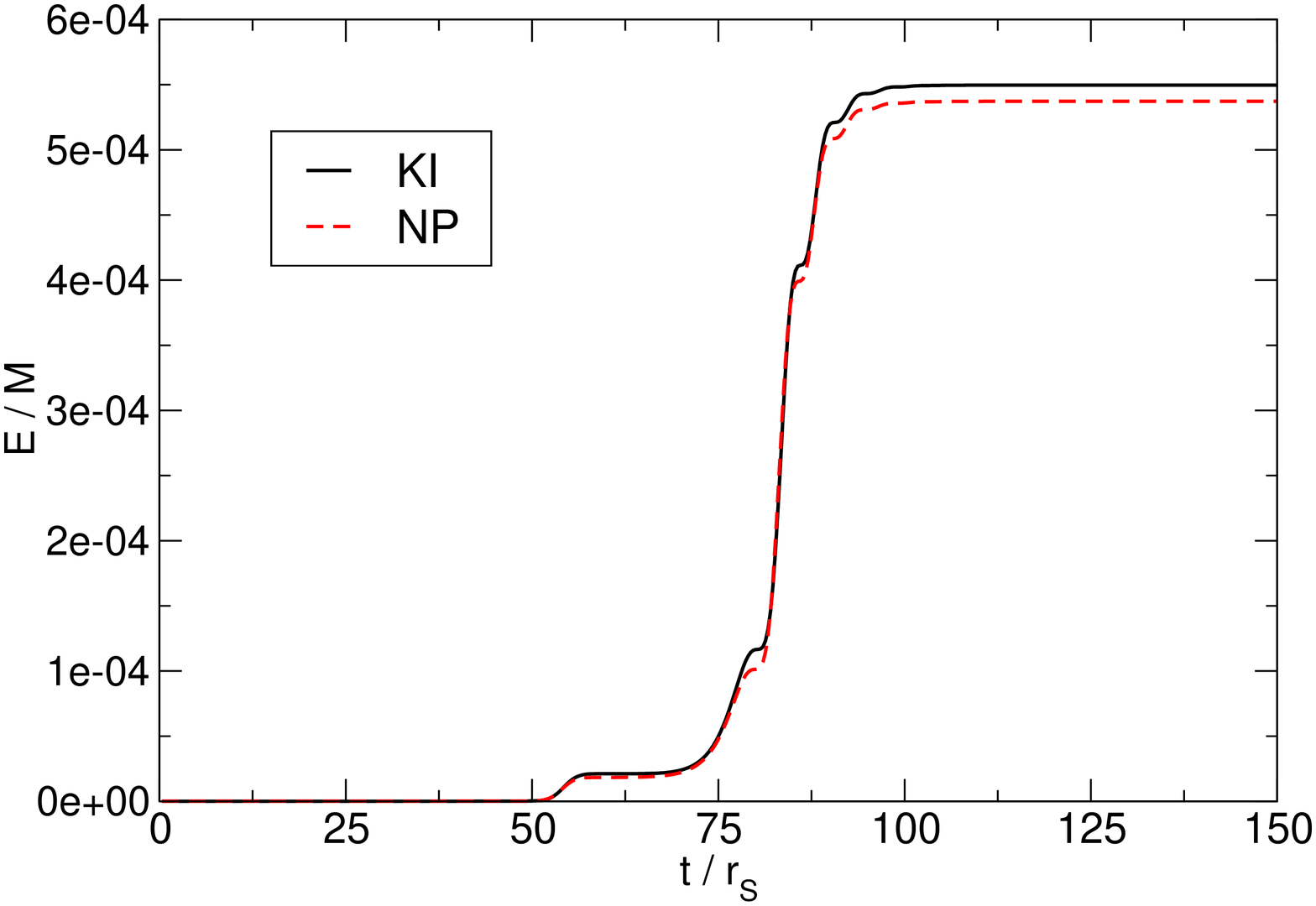} 
\end{tabular}
\end{center}
\caption{\label{fig:energyflux} (Color online) Energy flux (left panel)
  and radiated energy (right panel) for the $l=2$ mode extracted at $R_{\rm ex} = 50r_S$
  from the KI wave function $\Phi_{,t}$  (black solid curve)
  and the Newman Penrose scalar $\Psi_4$ (red dashed curve).
}
\end{figure*}

Once the KI function $\Phi_{,t}$ is known, the energy flux
can be computed from Eq.~(\ref{eq:energyflux}). For comparison, we have 
also determined the flux from the outgoing Newman Penrose scalar $\Psi_4$ according to Eq.~(22) 
in Ref.~\cite{Witek:2010qc}. The flux and energy radiated in 
the $l=2$ multipole, obtained with the two methods at $R_{\rm ex}=50\,r_S$
is shown in Fig.~\ref{fig:energyflux} and demonstrates agreement within the numerical uncertainties
of about $4~\%$ for either result.
We obtain an integrated energy of $5.5\times10^{-4}~M$
and $5.3\times10^{-4}~M$, respectively, for the gravitational wave energy
radiated in $l=2$, where $M$ denotes the centre of mass energy.

The energy in the $l=2$ mode is known to contain
more than $99\%$ of the total radiated energy \cite{Sperhake:2006cy}.
Our analysis is compatible with this finding; while the energy in the $l=3$
mode is zero by symmetry, our result for the energy in the
$l=4$ mode obtained from the KI master function is three orders of magnitude smaller than that of the $l=2$
contribution.

\section{Head-on collision from rest in $D=5$}
\label{sec:5D}
Having tested the wave extraction formalism in four dimensions,
we now turn our attention to the new results obtained for head-on collisions of BHs in five dimensional space-times. As before, we consider nonspinning BH binaries initially at rest with coordinate separation $d$.
Note that in five space-time dimensions the Schwarzschild radius is 
related to the ADM mass $M$ via Eq.~\eqref{defadm},
\begin{equation}
r_S^2=\frac{8M}{3\pi}\,.\label{defadm5d}
\end{equation}
We therefore define the ``total'' Schwarzschild radius
$r_S$ such that $r_S^2=r_{S,1}^2 + r_{S,2}^2$. By using this definition,
$r_S$ has physical dimension of length and provides a suitable
unit for measuring both, results and grid setup.

As summarized in Table \ref{tab:setup}, we consider a
sequence of BH binaries with initial coordinate
separation ranging from $d = 3.17\,r_{S}$ to $d = 10.37\,r_{S}$.
The table further lists the proper separation $L$ along the line of sight
between the holes and the grid configurations used for the individual
simulations.

\subsection{Tests on the numerical coordinates}\label{sec:D5coordinatecheck}
%
\begin{figure*}
\begin{center}
\begin{tabular}{cc}
\includegraphics[width=0.45\textwidth]{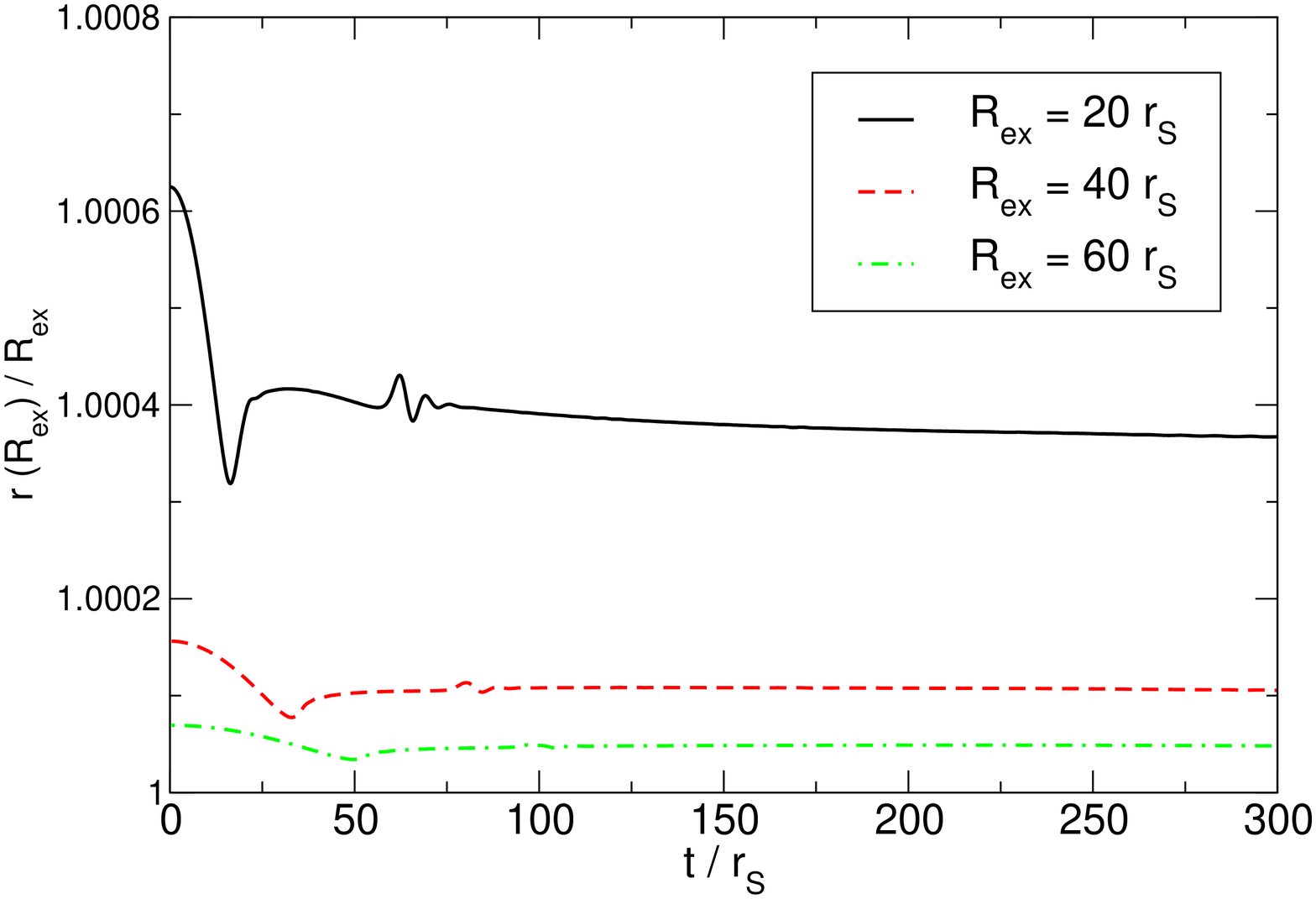} &
\includegraphics[width=0.45\textwidth]{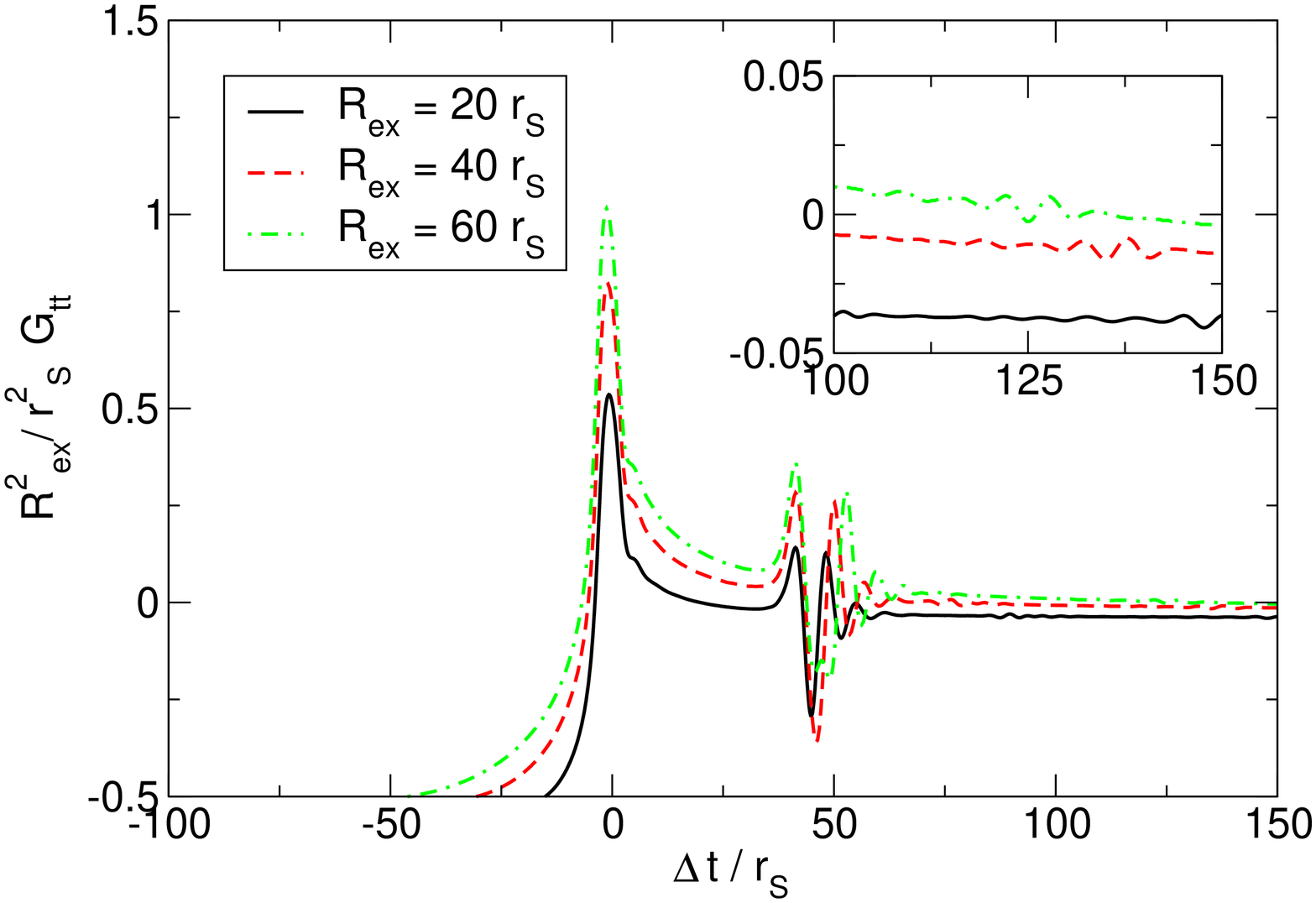} 
\end{tabular}
\end{center}
\caption{\label{fig:D5coordinates} (Color online)
Left panel:
     Time evolution of the areal radius $r$ in units of the extraction
     radius averaged over coordinate spheres at
     $R_{\rm ex}=20~r_S$ (black solid), $40~r_S$ (red dashed)
     and $60~r_S$ (green dash-dotted curve).
Right panel:
     Deviation of the metric component $R_{\rm ex}^2/r_S^2\mathcal{G}_{tt}$
     calculated from Eq.~(\ref{eq:testgtt}) at the same
     extraction radii and shifted in time to account for differences
     in the propagation time of the wave signal.
}
\end{figure*}

In order to verify the assumptions underlying our formalism,
we have analysed the coordinate system in analogy to Sec.~\ref{sec:coordtestD4}. First, we have evaluated the
averaged areal radius on extraction spheres of constant coordinate
radius. 

The result shown in the left panel of Fig.~\ref{fig:D5coordinates} demonstrates that the coordinate and areal
radius agree within about 1 part in $10^4$ for $R_{\rm ex}\ge40~r_S$.
The Tangherlini coordinate $r$ equals by construction the areal
radius and our approximation of setting $r\approx R_{\rm ex}$ in the
wave extraction zone is satisfied with high precision.

Second, we evaluate the deviation of the metric components according to
Eqs.~(\ref{eq:testgtt})-(\ref{eq:testgRR}). From the discussion in Sec.~\ref{sec:coordtestD4} we
expect $\mathcal{G}_{tt}\sim r^2 /r_S^2$ in $D=5$. Our results in the right panel of Fig.~\ref{fig:D5coordinates} confirm 
this expectation and demonstrate that our space-time is indeed perturbatively
close to that of a Tangherlini metric at sufficient distances from
the black holes; deviations in $\mathcal{G}_{tt}$ are well below 1
part in $10^3$ at $R_{\rm ex}=60~r_S$.
Furthermore, we can 
estimate the crudeness of the approximation $r^2_S = r^2_{S,1} + r^2_{S,2}$ for the Schwarzschild radius of the final BH:
as shown in the right panel of Fig.~\ref{fig:D5coordinates}, at late times
$|R_{ex}^2/r_S^2\mathcal{G}_{tt}|\sim 0.01$; 
this value gives an upper bound on the radiated energy.

For the third test, we recall that our higher dimensional implementation
does not employ the full isometry group of the $S^2$ sphere in
$D=5$ dimensions and axial symmetry manifests itself instead in the
conditions (\ref{checkax}) on the metric components and the scalar field.
We find these conditions to be satisfied within 1 part in $10^8$ and 1 part in $10^{16}$, respectively, in our numerical
simulations which thus represent axially symmetric configurations with high precision.

\subsection{Newtonian collision time}
\begin{figure}
\begin{center}
\begin{tabular}{cc}
\includegraphics[width=0.45\textwidth]{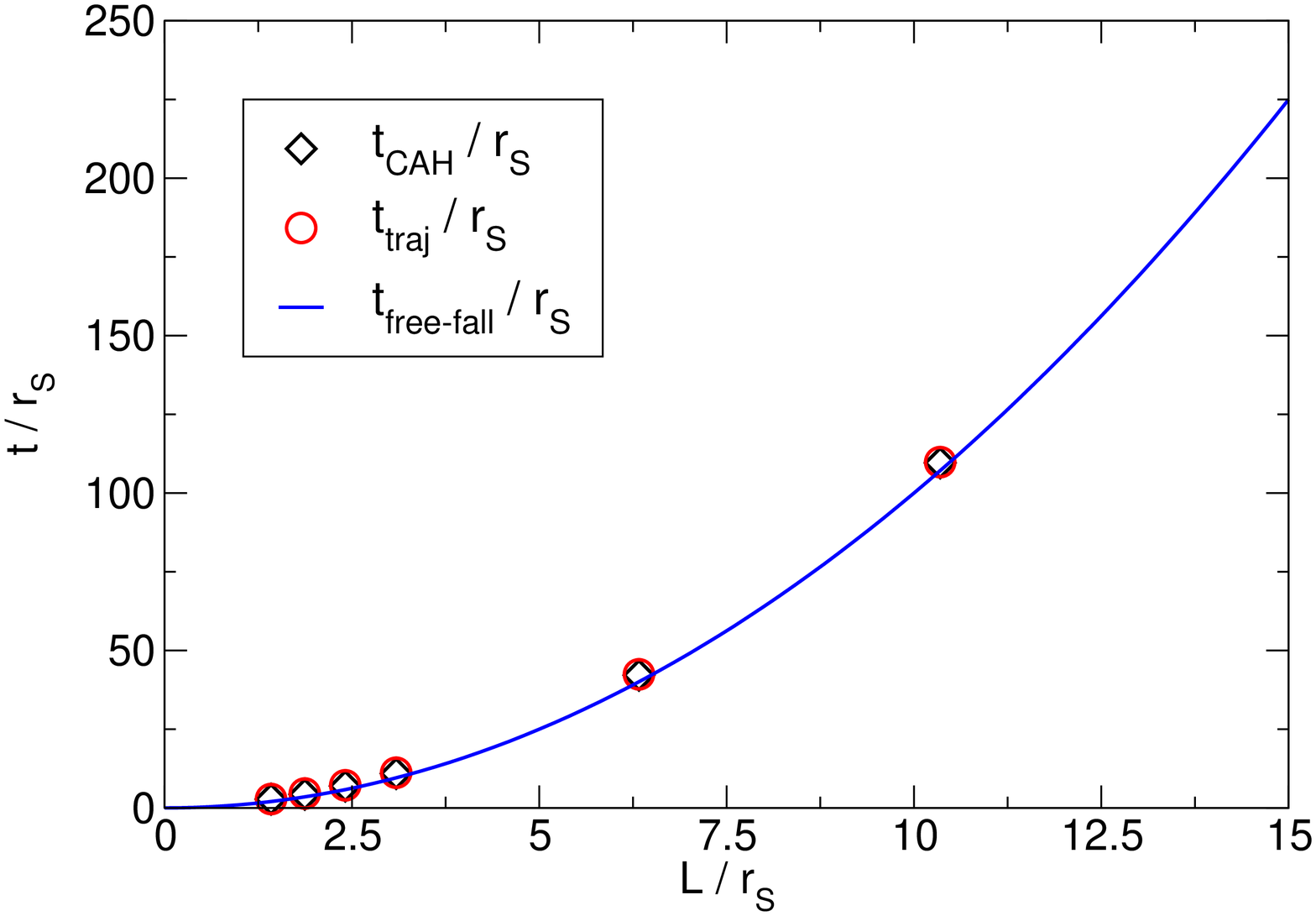} 
\end{tabular}
\end{center}
\caption{\label{fig:D5_NewtCollTime} 
(Color online) Estimates for the time it takes for two equal-mass BHs to
collide in $D=5$.  The first estimate is given by the time $t_{\text{CAH}}$
elapsed until a single common apparent horizon engulfs both BHs
(diamonds), the second estimate is obtained by using the trajectory of
the BHs, i.e., the time $t_{\text{traj}}$ at which their separation
has decreased below the Schwarzschild radius (circles). Finally, these
numerical results are compared against a simple Newtonian estimate,
given by Eq.~(\ref{eq:5dtime}) (blue solid line).
}
\end{figure}

An estimate of the time at which the holes ``collide'', can be obtained
by considering a Newtonian approximation to the kinematics of two point
particles in $D=5$. In the weak-field regime, Einstein's equations
reduce to ``Newton's law'' $a=-\nabla {\cal B}({\bf x})$, with $h_{00}=-2{\cal
B}({\bf  x})=r_S^{D-3}/2r^{D-3}$.
The Newtonian time it takes for two point-masses (with Schwarzschild
parameters $r_{S,1}$ and $r_{S,2}$) to collide from rest with initial
distance $L$ in $D$ dimensions is then given by
\begin{equation}
\label{eq:newt-time}
\frac{t_{\text{free-fall}}}{r_S} = \frac{\mathcal{I}}{D-3}\left(\frac{L}{r_S}\right)^{\frac{D-1}{2}}\,,
\end{equation}
where $r_S^{D-3} =r_{S,1}^{D-3} + r_{S,2}^{D-3}$ and
\begin{equation}
\label{eq:int}
\mathcal{I} = \int_0^1 \sqrt{\frac{z^{\frac{5-D}{D-3}}}{1-z}} dz
= \sqrt{\pi} \frac{\Gamma(\frac{1}{2} 
  + \frac{1}{D-3})}{\Gamma(1 + \frac{1}{D-3})}\,.
\end{equation}
For $D=4$, one recovers the standard result
$t_{\text{free-fall}} = \pi \sqrt{L^3/r_S^3}r_S\,,$ whereas for $D=5$ we get
\begin{equation}
t_{\text{free-fall}}=\left(L/r_S\right)^2r_S\,.\label{eq:5dtime}
\end{equation}
In general relativity, BH trajectories
and merger times are intrinsically observer dependent quantities.
For our comparison with Newtonian estimates we have chosen relativistic
trajectories as viewed by observers adapted to the numerical coordinate
system. While the lack of fundamentally gauge invariant analogues in
general relativity prevents us from deriving rigorous conclusions,
we believe such a comparison to serve the intuitive interpretation
of results obtained within the ``moving puncture'' gauge.
Bearing in mind these caveats, we plot in Fig.~\ref{fig:D5_NewtCollTime} 
the analytical estimate of the Newtonian time of collision, together
with the numerically computed time of formation of a common apparent horizon.
Also shown in Fig.~\ref{fig:D5_NewtCollTime} is the time at
which the separation between the individual hole's puncture trajectory decreases below
the Schwarzschild parameter $r_S$. 
The remarkable agreement provides yet another example of how well
numerically successful gauge conditions appear to be adapted to the
black hole kinematics. It is beyond the scope of this paper to
investigate whether this is coincidental or whether such agreement is
necessary or at least helpful for gauge conditions to ensure numerical
stability. Suffice it to say at this stage that
similar conclusions were reached by Anninos {\em et al.} \cite{Anninos:1994gp} and
Lovelace {\em et al.} \cite{Lovelace:2009dg} in similar four dimensional scenarios.

\subsection{Waveforms}\label{sec:D5waveforms}
%
\begin{figure*}
\begin{center}
\begin{tabular}{cc}
\includegraphics[width=0.45\textwidth]{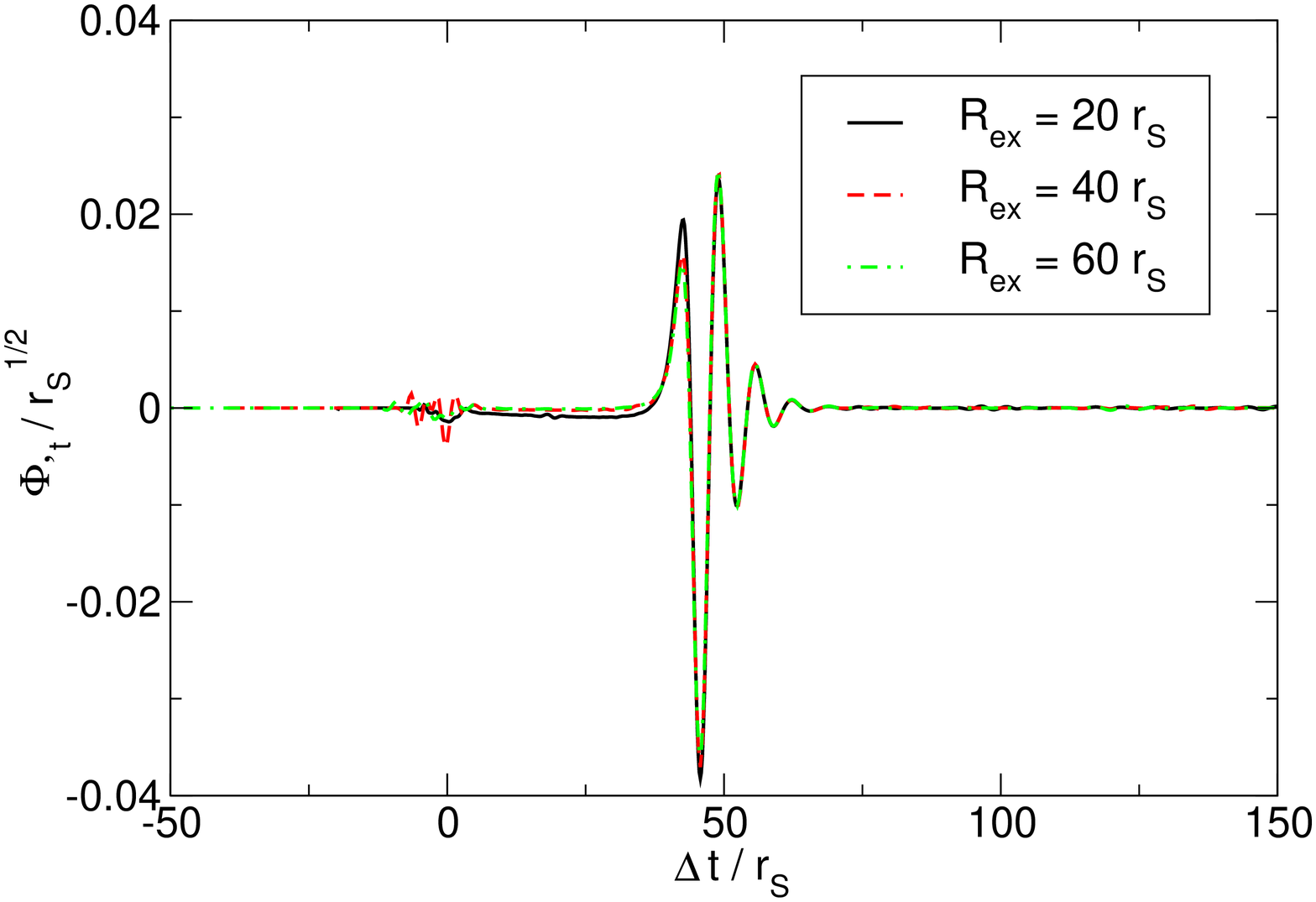} &
\includegraphics[width=0.45\textwidth]{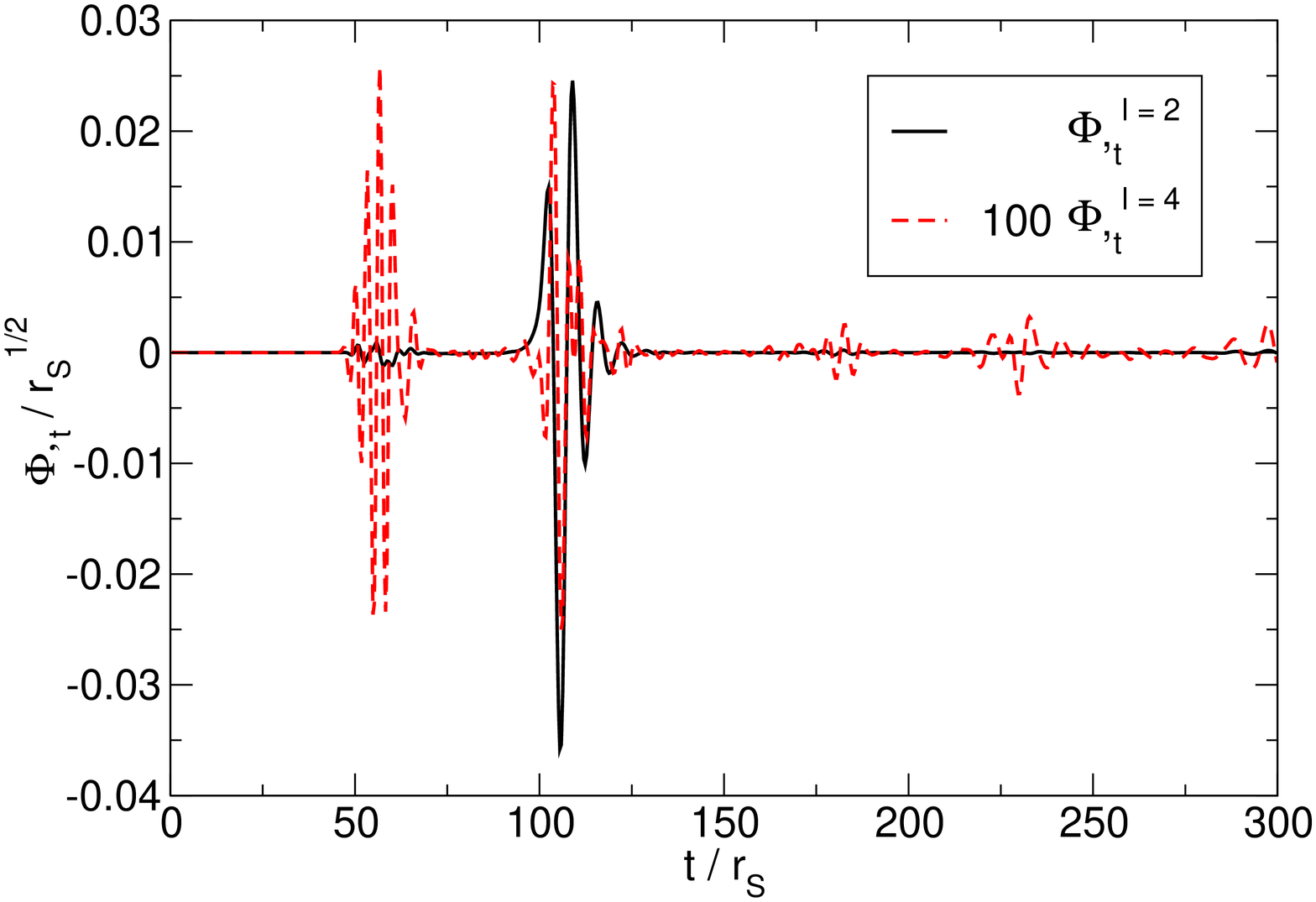} 
\end{tabular}
\end{center}
\caption{\label{fig:D5_Phir_hf84} (Color online)
Left panel: The $l=2$ component of the KI waveform for model HD5e$_f$  
extracted at radii $R_{\rm ex}/r_{S} = 20, 40$ and $60$ and shifted in time by 
$R_{\rm ex}/r_S$. Right panel: The $l=2$ and $l=4$ mode of the KI function for the same 
simulation, extracted at $R_{\rm ex}/r_S = 60$. For clarity, the $l=4$
component has been re-scaled by a factor of $100$.
}
\end{figure*}
\begin{figure}
\begin{center}
\begin{tabular}{cc}
\includegraphics[width=0.45\textwidth]{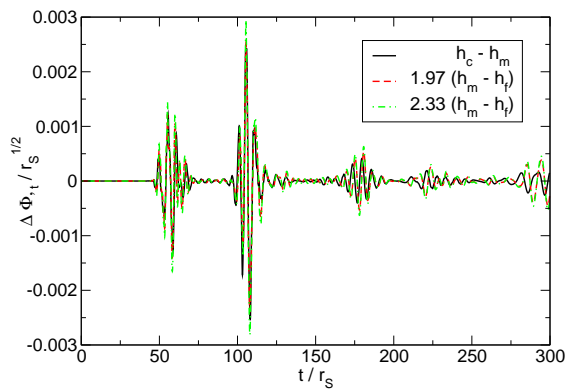} 
\end{tabular}
\end{center}
\caption{\label{fig:D5_convergence} (Color online) Convergence analysis
  of the $l=2$ component of the KI
  function generated by model HD5e  extracted at $R_{\rm ex}
  = 60r_{S}$.  The difference between the medium and high resolution
  waveforms has been amplified by the factors $1.97$ (red dashed line)
  and $2.33$ (green dashed-dotted line) indicating third and fourth
  order convergence. }
\end{figure}
We now discuss in detail the gravitational wave signal generated
by the head-on collision of two BHs in five dimensions.
For this purpose, we plot in Fig.~\ref{fig:D5_Phir_hf84}
the $l=2$ multipole of the KI function $\Phi_{,t}$ for
model HD5e$_f$ obtained at different extraction radii.
Qualitatively, the signal looks similar to that
shown in the left panel of Fig.~\ref{fig:comparison1} for $D=4$. A small spurious wavepulse
due to the initial data construction is visible at $\Delta t \approx 0$.
This so-called ``junk radiation'' increases in magnitude if the
simulation starts with smaller initial separation of the holes.
We return to this issue further below, when we study the dependence of the
gravitational radiation on the initial BH separation.
The physical part of the waveform is dominated by the merger signal
around $\Delta t=50~r_S$, followed by the (exponentially damped) ringdown,
whereas the infall of the holes before $\Delta t=40~r_S$ does not produce
a significant amount of gravitational waves.
Comparison of the waveforms extracted at different radii
demonstrates excellent agreement, in particular for those extracted at $R_{\rm ex}=40~r_S$ and
$60~r_S$. Extrapolation of the radiated energy
to infinite extraction radius yield a relative error of 5~\% at $R_{\rm ex}=60~r_S$,
indicating that such radii are adequate for the analysis presented in this work.

Due to symmetry, no gravitational waves are emitted in the $l=3$
multipole, so that $l=4$ represents the second strongest contribution
to the wave signal. As demonstrated in the right panel of
Fig.~\ref{fig:comparison1}, however, its amplitude is two orders
of magnitude below that of the quadrupole.

A convergence analysis also using the lower resolution simulations
of models HD5e$_c$ and HD5e$_m$ is shown in Fig.~\ref{fig:D5_convergence}
and demonstrates overall convergence of third to fourth order, consistent with
the numerical implementation.
From this analysis we obtain a conservative estimate of
about $4\%$ for the discretization error in the waveform.

\begin{figure}
\begin{center}
\begin{tabular}{cc}
\includegraphics[width=0.45\textwidth]{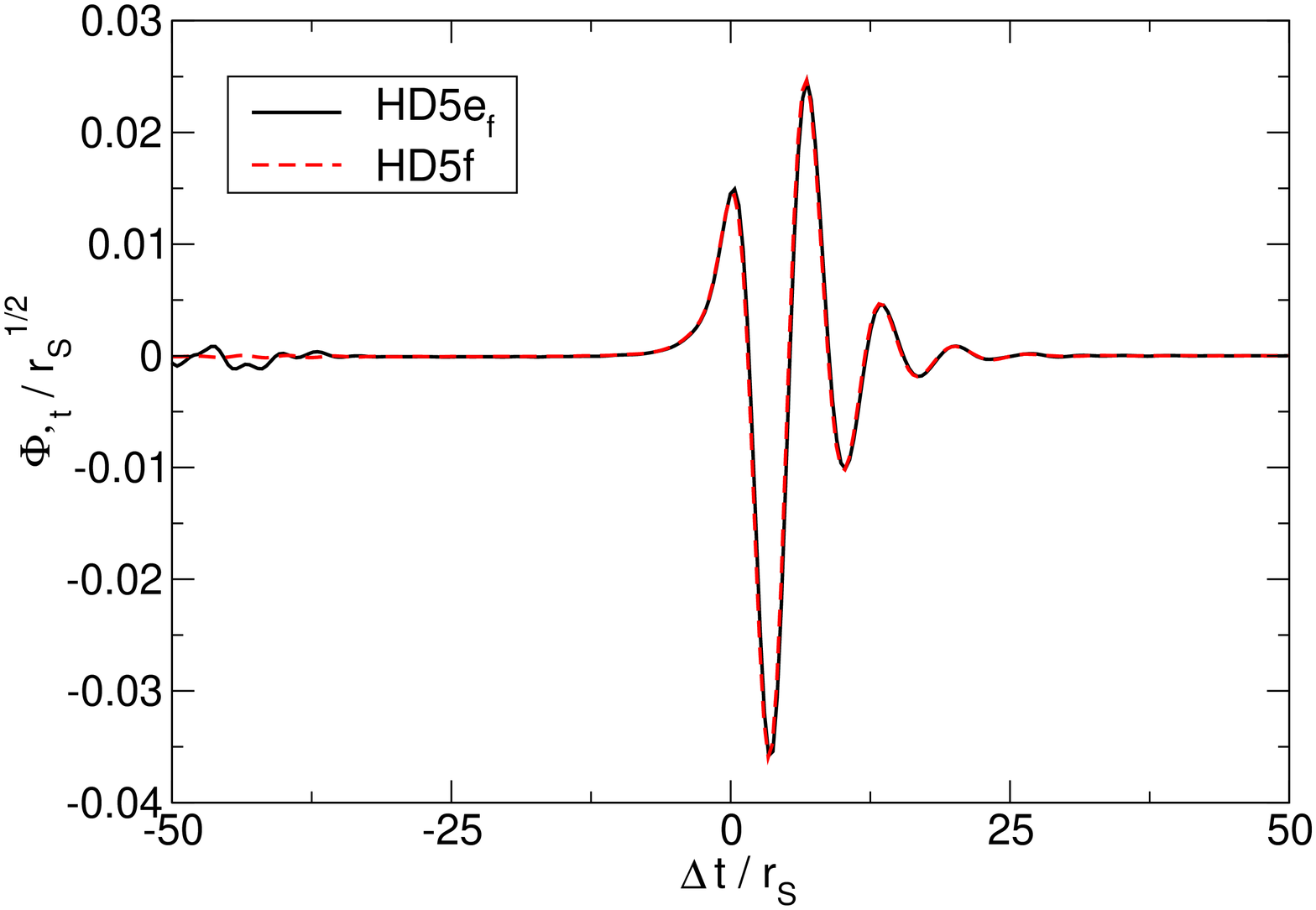} 
\end{tabular}
\end{center}
\caption{\label{fig:D5_Phit_models} (Color online)
The $l=2$ components of the KI function as generated by
a head-on collision of BHs with initial (coordinate) distance
$d=6.37\,r_S$ (black solid line) and $d=10.37\,r_S$ (red dashed line).
The wave functions have been shifted in time such that the formation of
a common apparent horizon corresponds to $\Delta t=0$ and taking into account
the time it takes for the waves to propagate up to the extraction radius
$R_{\rm ex} = 60\,r_{S}$.}
\end{figure}

In practice, numerical simulations will always start with a
finite separation of the two black holes. In order to assess how
accurately we are thus able to approximate an infall from infinity, we
have varied the initial separation for models HD5a to HD5f as
summarized in Table \ref{tab:setup}. For small $d$ we observe two
effects which make the physical interpretation of models HD5a$-$HD5c
difficult. First, the amplitude of the spurious initial
radiation increases and second, the shorter infall time causes an
overlap of this spurious radiation with the merger signal.
As demonstrated in Fig.~\ref{fig:D5_Phit_models} for models HD5e and HD5f,
however, we can safely neglect the spurious radiation as well
as the impact of a final initial separation, provided
we use a sufficiently large initial distance $d \gtrsim 6~r_S$
of the BH binary. Here, we
compare the radiation emitted during the head-on collision 
of BHs starting from rest with
initial separations $6.37~r_S$ and $10.37~r_S$.
The waveforms have been shifted in time by the extraction
radius $R_{\rm ex} = 60\,r_{S}$ and such that the formation of a common
apparent horizon occurs at $\Delta t=0$. The merger signal
starting around $\Delta t=0$ shows excellent agreement for the
two configurations and is not affected by the spurious signal
visible for HD5e at $\Delta t\approx -50~r_S$.

We conclude this discussion with two aspects of the post-merger
part of the gravitational radiation, the ringdown and the possibility
of GW tails. After formation of a common horizon,
the waveform is dominated by an exponentially damped
sinusoid, as the merged hole {\em rings down} into a stationary state.
By fitting our results with a exponentially damped sinusoid,
we obtain a characteristic frequency
\begin{equation}
r_{S}\,\omega=0.955\pm 0.005-{\rm i}\,(0.255 \pm 0.005)\,.\label{qnm}
\end{equation}
This value is in excellent agreement with perturbative calculations, which
predict a lowest quasinormal frequency $r_{S}\,
\omega=0.9477-{\rm i}\,0.2561$ for $l=2$
\cite{Cardoso:2003qd,Yoshino:2005ps,Berti:2009kk}.

A well known feature in gravitational waveforms generated in
BH space-times with $D=4$ as well as $D>4$ are the so-called
{\it power-law tails} 
\cite{DeWitt:1960fc,Price:1971fb,Ching:1995tj, Cardoso:2003jf}.
In odd dimensional space-times an additional, different kind of late-time
power tails
arises, which does not depend on the presence of a BH. These
are due to a peculiar behavior of the wave-propagation
in {\it flat} odd dimensional space-times
because the Green's function has support inside the entire
light-cone \cite{Cardoso:2003jf}.
We have attempted to identify
such power-law tails in our signal at late-times,
by subtracting a best-fit ringdown waveform. Unfortunately, we
cannot, at this stage, report
any evidence of such a power-law in our results, most likely because the
low amplitude tails are buried in numerical noise.

%
\subsection{Radiated energy}\label{sec:D5raden}
%
\begin{figure}
\begin{center}
\begin{tabular}{c}
\includegraphics[width=0.45\textwidth]{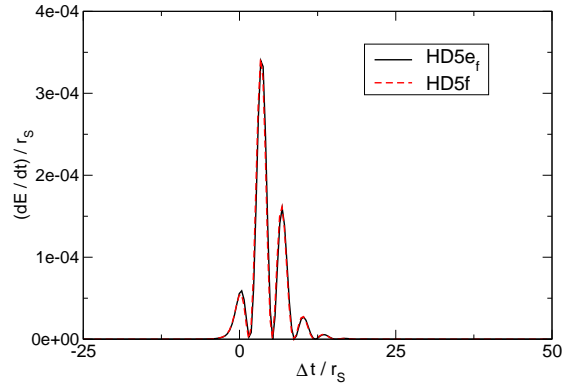} 
\end{tabular}
\end{center}
\caption{\label{fig:D5energy} 
(Color online) Energy flux in the $l=2$ component of the KI 
wave function   $\Phi_{,t}$, extracted at $R_{\rm ex} = 60\,r_{S}$, for models 
HD5e$_f$ (black solid line) and HD5f (red dashed line)
in Table~\ref{tab:setup}. The fluxes have been shifted in time by 
the extraction radius $R_{\rm ex} = 60\,r_{S}$ and the time $t_{\text{CAH}}$ 
at which the common apparent horizon forms.
}
\end{figure}
Comparison of Figs.~\ref{fig:Phir} and \ref{fig:D5_Phit_models}
for the GW quadrupole in $D=4$ and $D=5$ shows a larger wave amplitude
in the five dimensional case and thus indicates that this case
may radiate more energy. We now investigate this question quantitatively
by calculating the energy flux from the KI master function via
Eq.~(\ref{eq:energyflux}). The fluxes thus obtained for
the $l=2$ multipole of models HD5e$_f$ and HD5f
in Table~\ref{tab:setup}, extracted at $R_{\rm ex} = 60\,r_{S}$,
are shown in Fig.~\ref{fig:D5energy}.
As in the case of the KI master function in
Fig.~\ref{fig:D5_Phit_models}, we see no significant variation of the
flux for the two different initial separations.
The flux reaches a maximum value of $dE/dt\sim 3.4\times 10^{-4}\,r_S$,
and is then dominated by the ringdown flux. The energy flux from
the $l=4$ mode is typically four orders of magnitude smaller; this is consistent with the factor of 100 difference of the corresponding
wave multipoles observed in Fig.~\ref{fig:D5_Phir_hf84}, and the quadratic
dependence of the flux on the wave amplitude.

\begin{figure*}
\begin{center}
\begin{tabular}{cc}
\includegraphics[width=0.45\textwidth]{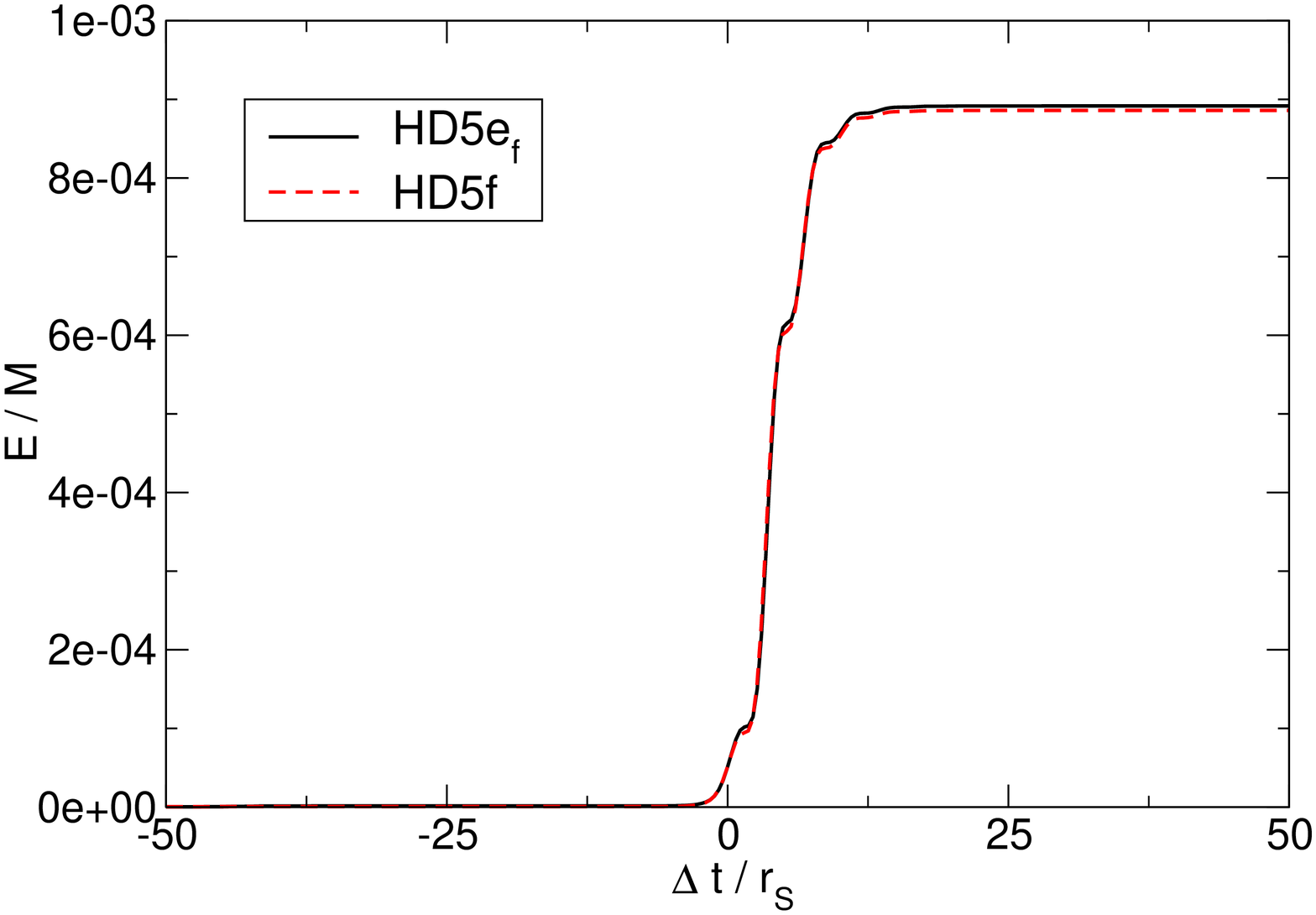}&
\includegraphics[width=0.45\textwidth]{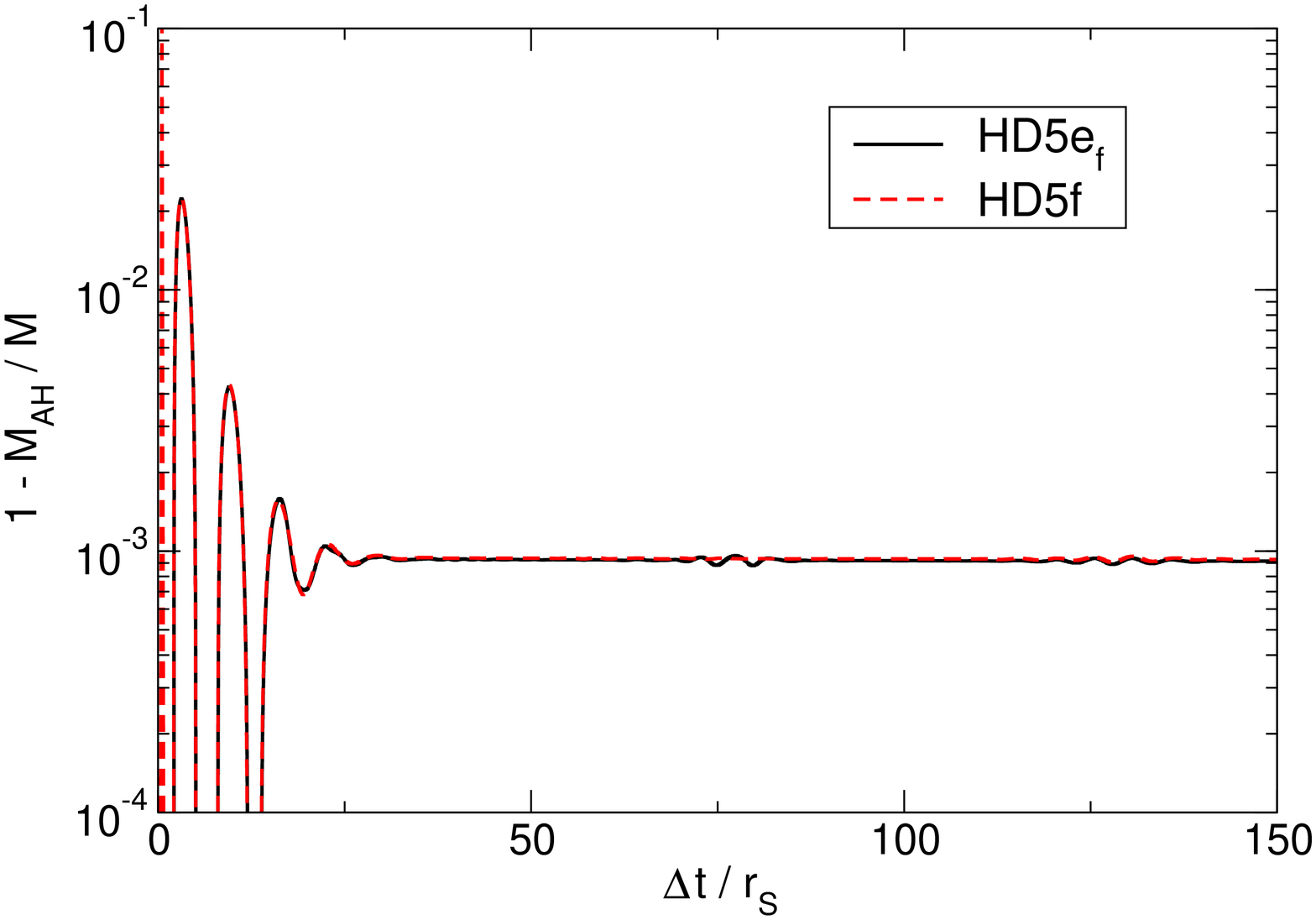}
\end{tabular}
\end{center}
\caption{\label{fig:totalenergy} (Color online)
Left panel: Fraction of the centre of mass energy, $E_{rad}/M$, radiated
in the $l=2$ mode of the KI function shifted in time
such that the origin of the
time axis corresponds to the formation of a common apparent horizon.  %
Right panel: Fraction of the centre of mass energy $1-M_{AH}/M$ radiated
during the collision, estimated using apparent horizon information.
The oscillations in this diagnostic quantity have a frequency comparable
to the $l=2$ quasinormal mode frequency.
}
\end{figure*}
The total integrated energy emitted throughout the head-on collision
is presented in the left panel of Fig.~\ref{fig:totalenergy}. We find that a fraction of
$E_{rad}/M = (8.9\pm0.6)\times 10^{-4}$ of the centre of mass energy
is emitted in the form of gravitational radiation. 
We have verified for these models that
the amount of energy contained in the spurious radiation is about
three orders of magnitude smaller than in the physical merger signal.

An independent estimate for the radiated energy can be obtained from the apparent
horizon area $A_4$ in the effective four dimensional space-time
by using the spherical symmetry of the post-merger remnant hole. Energy balance then implies that the
energy $E$ radiated in the form of GWs is given by
\begin{equation}
  \frac{E}{M} = 1-\frac{M_{AH}}{M} =
1-\frac{A_4}{4\pi r_S^2}\,,
\end{equation}
where $M_{AH}$ is the apparent horizon mass. 
The estimate $E/M$ is shown in Fig.~\ref{fig:totalenergy} 
and reveals a behavior qualitatively similar to 
a damped sinusoid with constant offset. Indeed, by using
a least square fit, we obtain a complex frequency
$r_{S}\,\omega\sim0.97-{i}\,0.29$, again similar to the fundamental $l=2$ quasinormal mode
frequency (see discussion around Eq.~(\ref{qnm})).
At late times, $1-M_{AH}/M$ asymptotes
to $1-M_{AH}/M\sim \left(9.3\pm 0.8\right)\times10^{-4}$
which agrees very well with the GW estimate, within the numerical uncertainties.
\section{Discussion\label{final}}
In this paper we have developed a formalism to extract gravitational
radiation observables from numerical simulations of head-on collisions of BHs
in $D$ dimensions. Moreover, we have performed such simulations in
$D=4,5$. 
The $D=4$ case serves as a test of our formalism
and demonstrates consistency of our results with the literature.
The $D=5$ case is entirely new. Besides obtaining the corresponding waveforms, we
have shown that the total energy released in the form of gravitational
waves  is approximately $(0.089\pm 0.006)\%$ of the initial centre of mass energy
of the system, for a head-on collision of two BHs starting from rest
at very large distances. As a comparison, the analogous process in
$D=4$ releases a slightly smaller quantity: $(0.055\pm 0.006)\%$.
We summarize the main results for head-on collisions of two BHs starting from rest
in four and five space-time dimensions in Table \ref{tab:summary}.
\begin{table*}
\caption{\label{tab:summary} Main results for head-on collisions
in $D=4$ and $5$ dimensions. We list the ring down frequency $\omega$,
the total energy radiated in gravitational waves, the upper bound
$E^{\rm area}$ on the radiated energy obtained from Hawking's area theorem
and the fractional energy in the $l=4$ multipole relative to the
quadrupole radiation.}
\begin{tabular}{ccccc}
\hline
$D$ & $r_{S}\,\omega (l=2)$ &$E^{\rm rad}/M(\%)$ &$E^{\rm area}/M(\%)$&$E^{\rm rad}_{l=4}/E^{\rm rad}_{l=2}$\\ \hline
$4$ & $0.7473-{\rm i}\,0.1779$                              &$0.055$                   & $29.3$ &$<10^{-3}$ \\
$5$ & $0.9477-{\rm i}\,0.2561$                              &$0.089$                   & $20.6$ &$<10^{-4}$ \\
\hline
\end{tabular}
\end{table*}

We have further performed a variety of tests of the wave extraction
formalism. Besides testing the proximity of the
numerical coordinate system to the Tangherlini background space-time,
we have demonstrated good agreement between the radiated energy
as derived directly from the KI master function with the values
obtained from the horizon area of the post-merger remnant hole.
Finally, the ringdown part of the waveform yields a quasinormal mode
frequency in excellent agreement with predictions from BH
perturbation theory.

The radiative efficiency $E^{\rm rad}/M$ in Table \ref{tab:summary}
shows that head-on collisions starting from rest in five dimensions
generate about 1.6 times as much GW energy as their four dimensional
counterparts. It will be very interesting to investigate to what
extent this observation holds for wider classes of BH collisions.
We can compare the radiation efficiency with the
upper limit derived by Hawking \cite{Hawking:1971} from the
requirement that the horizon area must not decrease in the collision.
This leads to the {\em area bound}
%
$\frac{E^{\rm area}}{M} \leq 1-2^{-\frac{1}{D-2}}\,.$
%
Evidently, this bound decreases with dimensionality, 
while in the present computation it increases when going from $D=4$ to $D=5$.
As also shown in the table, the generation
of GWs in head-on collisions starting from rest is about 3 orders
of magnitude below this bound. In four dimensions it has already been demonstrated that there exist
more violent processes which release more radiation than the
head-on collisions considered in this work
\cite{Sperhake:2008ga, Shibata:2008rq, Sperhake:2009jz}.

In the context of this work, it would be particularly interesting to
compute the gravitational radiation emitted when a point particle falls
into a higher dimensional BH (the four dimensional calculation is done
in the classic work by Davis et al \cite{Davis:1971gg}).  This analysis
can be done by linearizing Einstein's equations. While such an analysis
was done for infall at high energies \cite{Cardoso:2002ay,Berti:2003si},
it has not been done for infalls from rest.  The four dimensional case
shows that by scaling the point-particle results properly with the reduced
mass, one gets surprisingly good agreement with full nonlinear studies
\cite{Anninos:1993zj}. An obvious question is whether such an agreement
extends to generic number of space-time dimensions. Investigations with
a similar purpose, but using a different technique, were carried out
in Refs.~\cite{Araujo:1995vb,Moreschi:1996ds}.

The numbers reported here for the total energy loss in gravitational waves
should increase significantly in high energy collisions, 
which are the most relevant scenarios for the applications described in the
Introduction. Indeed, in the four dimensional case, it is known that 
ultra-relativistic head-on collisions of equal mass nonrotating BHs release up
to 14\% of the initial centre of mass energy into gravitational radiation
\cite{Sperhake:2008ga}. The analogous number in higher dimensions is
as yet unknown and  will be subject of the next stages of our
research programme. Preliminary results by Gal'tsov et al \cite{Gal'tsov:2009zi,Gal'tsov:2010me} strongly suggest enhancement of radiation emission for higher dimensions,
in agreement with the $D=5$ results shown here. Even more energy may be released in high energy collisions with nonvanishing impact parameter. In \cite{Shibata:2008rq,Sperhake:2009jz} it was shown that this number can be as large as 35\% in $D=4$.
The formalism developed in Paper I allows, in principle, the study of analogous processes in $D\ge 6$. 
We hope to be able to report on these results in the near future. 

\begin{acknowledgments}

. A.N., H.W. and M.Z. would like to express their gratitude to the 
  Department of Physics and Astronomy of the University of Mississippi
  for their hospitality during the last stages of the work. 
  We thank Emanuele Berti and Marco Cavagli\`a for useful conversations and suggestions. 
  M.Z. and H.W. are funded by FCT through grants SFRH/BD/43558/2008 and SFRH/BD/46061/2008. 
  A.N. is funded by FCT through grant SFRH/BPD/47955/2008. This work was supported by the 
  {\it DyBHo--256667} ERC Starting Grant, by Funda\c c\~ao Calouste Gulbenkian, by FCT - Portugal 
  through projects PTDC/FIS/098025/2008, PTDC/FIS/098032/2008, PTDC/CTE-AST/098034/2008, 
  CERN/FP/109306/2009, CERN/FP/109290/2009, by the Ram\'on y Cajal Programme of the Ministry of Education 
  and Science of Spain, NSF grants PHY-0601459, PHY-0652995 and the Fairchild Foundation to Caltech, 
  as well as NSF grant PHY-0900735. This research was supported by an allocation through the 
  TeraGrid Advanced Support Program under grant PHY-090003 and an allocation by the 
  Centro de Supercomputaci{\'o}n de Galicia (CESGA) under project ICTS-2009-40. 
  Computations were performed on the TeraGrid clusters TACC Ranger and NICS Kraken, 
  at Magerit in Madrid, Finis Terrae and the Milipeia cluster in Coimbra. 
  The authors thankfully acknowledge the computer resources, technical expertise and 
  assistance provided by the Barcelona Supercomputing 
  Centre---Centro Nacional de Supercomputaci\'on.
\end{acknowledgments}

\appendix

\section{Coordinate transformation}\label{app:coordtrafo}
In order to extract gravitational radiation using the KI formalism one has to perform a coordinate transformation from 
Cartesian coordinates, which are used during the numerical evolution, to those adapted
for wave extraction.
The physical 3-metric $\gamma_{ij}$, the lapse function $\alpha$ and the shift vector $\beta^i$ computed on our
Cartesian grid are interpolated onto a Cartesian patch. In terms of these quantities we compute the 4-metric
$g_{\mu\nu}$ in Cartesian coordinates according to Eq.~\eqref{totalmetric}:
\begin{align}
g_{\mu\nu}dx^{\mu}dx^{\nu} =& (-\alpha^2 + \gamma_{ij} \beta^i\beta^j)dt^2
                             +\gamma_{ij}\beta^i dt dx^j \nonumber\\
                            &+ \gamma_{ij}\beta^j dt dx^i + \gamma_{ij} dx^i dx^j\,.
\end{align}
Then, we transform the 4-metric in Cartesian coordinates into spherical coordinates, defined by Eq.~\eqref{transfc}
\begin{align}
x&=
R\sin\bar\theta\cos\theta \, , \\
y&=R\sin\bar\theta\sin\theta \, ,\\
z&=R\cos\bar\theta \, , 
\end{align}\label{transfc1}
where $\bar\theta,\theta\in[0,\pi]$ and $R=\sqrt{x^2+y^2+z^2}$.
If we denote the metric in spherical coordinates by $g_{\mu\nu}^S$
and define $\rho \equiv \sqrt{x^2 + y^2}$,
the explicit form of the transformation is
\begin{align}
g^S_{tR} =& g_{tx} \sin\bar{\theta} \cos\theta 
           + g_{ty} \sin\bar{\theta} \sin\theta 
           + g_{tz}\cos\bar{\theta} \, , \\
g^S_{t\bar{\theta}} =& z (g_{tx} \cos\theta 
                      +g_{ty}\sin\theta ) 
                      -\rho g_{tz} \, ,\\
g^S_{t\theta} =&-y g_{tx} + x g_{ty}\, , \\
g^S_{RR} =& g_{xx} \sin^2\bar{\theta} \cos^2\theta 
           +2g_{xy} \sin^2\bar{\theta} \cos\theta \sin\theta \nonumber\\
          &+2g_{xz} \sin\bar{\theta} \cos\theta  \cos\bar{\theta}   
           +g_{yy} \sin^2\bar{\theta} \sin^2\theta \nonumber\\
          &+2 g_{yz}  \sin\bar{\theta} \sin\theta  \cos\bar{\theta} 
           +g_{zz}\cos^2\bar{\theta}  \, , \\
g^S_{R\bar{\theta}} =& z ( g_{xx} \sin\bar{\theta} \cos^2\theta   
                      +2g_{xy} \sin\bar{\theta} \cos\theta  \sin\theta \nonumber\\ 
                     &+g_{yy} \sin\bar{\theta} \sin^2\theta 
                      +g_{xz} \cos\bar{\theta}  \cos\theta        
                      +g_{yz}\cos\bar{\theta}  \sin\theta   ) \nonumber\\
                     &-(x g_{xz} + y g_{yz} + z g_{zz})\sin\bar{\theta}\, , \\
g^S_{R\theta} =& ( -y  g_{xx} \sin\bar{\theta} \cos\theta  
                +x g_{xy}  \sin\bar{\theta} \cos\theta                
                -y g_{xy} \sin\bar{\theta} \sin\theta \nonumber\\
               &+x g_{yy} \sin\bar{\theta} \sin\theta                 
                -y g_{xz} \cos\bar{\theta} 
                +x g_{yz} \cos\bar{\theta}  )\, , \\
g^S_{\bar{\theta}\bar{\theta}} =& z^2 ( g_{xx} \cos^2\theta 
                                       +2 g_{xy} \cos\theta \sin\theta 
                                       +  g_{yy}\sin^2\theta )  \nonumber\\
                                &-2z (x g_{xz} + y  g_{yz}) 
                                 +\rho^2  g_{zz} \, ,\\
g^S_{\bar{\theta}\theta} =& z (-y g_{xx} \cos\theta 
                               +x g_{xy}    \cos\theta      
                               -y g_{xy} \sin\theta  \nonumber\\
                          &    +x g_{yy} \sin\theta )   
                           + \rho  (y g_{xz} - x  g_{yz}) \, , \\
g^S_{\theta\theta} =& R^2\sin^2\bar\theta ( g_{xx} \sin^2\theta 
                                           -2g_{xy}  \cos\theta  \sin\theta  
                                           +g_{yy} \cos^2\theta  ) \, .
\end{align}
Henceforth, we will drop the superscript $S$ and use $g_{\mu\nu}$ for the metric 
in spherical coordinates.

The areal radius $r$ is related to $R$ by a reparametrization $R=R(r)$, given by Eq.~\eqref{reparam}, which depends on
the components $g_{\bar\theta\bar\theta}$, $g_{\theta\theta}$ only. As shown in Section  \ref{numres},
we find that this reparametrization is nearly constant throughout our numerical simulations.
Therefore, the quantities $g_{rr},g_{tr}$, $g_{r\bar\theta},g_{r\theta}$ can be obtained from $g_{RR},g_{tR},g_{R\bar\theta},g_{R\theta}$
by a simple rescaling: because
\begin{equation}
\frac{dR}{dr}\simeq1 \,,
\end{equation}
we have $g_{rr}\simeq g_{RR}$, and similar relations hold for the other components.

\section{Harmonic expansion of axisymmetric tensors in $D$ dimensions}\label{appintegrals}
As discussed in Section \ref{implax}, scalar spherical harmonics in $D$ dimensions ${\cal
  S}_l(\bar\theta,\theta,\phi^1,\dots,\phi^{D-4})$ are solutions of Eq.~\eqref{eqscal}
\begin{equation}
\Box{\cal S}_l=\gamma^{\bar i\bar j}{\cal S}_{l\,:\bar i\bar j}=-k^2{\cal S}_l \, , 
\label{eqscal2}
\end{equation} 
with $k^2=l(l+D-3)$. Axisymmetric scalar spherical harmonics are functions of the coordinate $\bar\theta$ only, 
${\cal S}_l={\cal S}_l(\bar\theta)$. Therefore, Eq.~\eqref{eqscal2} becomes
\begin{equation}
\Box{\cal S}_l(\bar\theta)={\cal S}_{l\,,\bar\theta\bar\theta}+(D-3)\cot\bar\theta{\cal S}_{l\,,\bar\theta}
=-k^2{\cal S}_l \, ,\label{eqscal3}
\end{equation}
since
\begin{align}
{\cal S}_{l\,:\bar\theta\bar\theta}&={\cal S}_{l\,,\bar\theta\bar\theta} \, \\
{\cal S}_{l\,:\theta\theta}&=-\Gamma_{\theta\theta}^{\bar\theta}{\cal S}_{l\,,\bar\theta}=
\sin\bar\theta\cos\bar\theta{\cal S}_{l\,,\bar\theta} \, ,\\
{\cal S}_{l\,:\phi^1\phi^1}&=-\Gamma_{\phi^1\phi^1}^{\bar\theta}{\cal S}_{l\,,\bar\theta}=
\sin^2\theta \sin\bar\theta\cos\bar\theta{\cal S}_{l\,,\bar\theta}\, , 
\end{align}
etc. The quantities ${\cal S}_{l\,\bar i\bar j}$ defined in Eq.~\eqref{defharm} are then
\begin{align}
{\cal S}_{l\,\bar i\bar j} =& \frac{1}{k^2}{\cal S}_{l\,:\bar i\bar j}+\frac{1}{D-2}\gamma_{\bar i\bar j}{\cal S}_l 
  \nonumber\\
  =& \frac{1}{k^2(D-2)}\left((D-2){\cal S}_{l\,:\bar i\bar j}+k^2 \gamma_{\bar i\bar j}{\cal S}_l\right)
  \nonumber\\
  =& \frac{1}{k^2(D-2)}{\rm diag}\left((D-3){\cal W}_l, \right.\nonumber\\
   & \left. -\sin^2\bar\theta {\cal W}_l,-\sin^2\bar\theta\sin^2
     \theta {\cal W}_l,\dots \right)
\label{defharm2}
\end{align}
where 
\begin{equation}
{\cal W}_l(\bar\theta) = {\cal S}_{l\,,\bar\theta\bar\theta}-\cot\bar\theta{\cal S}_{l\,,\bar\theta}
=\sin\bar\theta\left(\frac{{\cal S}_{l\,,\bar\theta}}{\sin\bar\theta}\right)_{,\bar\theta}\,.\label{defww}
\end{equation}
Indeed, using Eq.~\eqref{eqscal3} one finds
\begin{align}
k^2(D-2) {\cal S}_{l\,\bar\theta\bar\theta} =& (D-2)  {\cal S}_{l\,,\bar\theta\bar\theta}+k^2{\cal S}_l \nonumber\\
  =&(D-3)( {\cal S}_{l\,,\bar\theta\bar\theta}-\cot\bar\theta{\cal S}_{l\,,\bar\theta})\, , \\
k^2(D-2) {\cal S}_{l\,\theta\theta} &=(D-2) {\cal S}_{l\,,\theta\theta}+k^2\sin^2\bar\theta{\cal S}_l \nonumber\\
  =& \sin^2\bar\theta((D-2)\cot\bar\theta{\cal S}_{l\,,\bar\theta}+k^2{\cal S}_l)\nonumber\\
  =& \sin^2\bar\theta(-{\cal S}_{l\,,\bar\theta\bar\theta}+\cot\bar\theta
     {\cal S}_{l\,,\bar\theta}) \, ,
\end{align}
and therefore
\begin{align}
{\cal S}_{l\,\bar\theta\bar\theta}=&\frac{D-3}{k^2(D-2)}{\cal W}_l \, ,\\
{\cal S}_{l\,\theta\theta}=&-\frac{\sin^2\bar\theta}{k^2(D-2)}{\cal W}_l \, ,
\end{align}
and likewise for the other components.

Axisymmetric scalar spherical harmonics, as discussed in Sec. \ref{implax}, can be written in terms of Gegenbauer
polynomials (cf. \eqref{gp}):
\begin{equation}
{\cal S}_l(\bar\theta)=(K^{lD})^{-1/2}C_l^{(D-3)/2}(\cos\bar\theta) \,.\label{defSC}
\end{equation} 
If we define
\begin{equation}
W_l(\cos\bar\theta)=C_{l\,,\bar\theta\bar\theta}^{(D-3)/2}(\cos\bar\theta) -\cot\bar\theta C_{l\,,\bar\theta}^{(D-3)/2}(\cos\bar\theta) \,,
\end{equation}
we have
\begin{equation}
{\cal W}_l(\bar\theta)=(K^{lD})^{-1/2}W_l^{(D-3)/2}(\cos\bar\theta) \,.\label{defWC}
\end{equation}

We impose the normalization \eqref{normalK} 
\begin{equation}
\int d\Omega^{D-2}{\cal S}_l{\cal S}_{l'}=\delta_{ll'}\,,~~~
\int d\Omega^{D-2}{\cal S}_{l\,,\bar\theta}{\cal S}_{l'\,,\bar\theta}=\delta_{ll'}k^2\,.\label{normalK1}
\end{equation}
Using
\begin{align}
\int_0^{\pi}d\bar\theta (\sin\bar\theta)^{D-3}
C_l^{(D-3)/2}(\cos{\bar\theta}) C_{l'}^{(D-3)/2}(\cos{\bar\theta}) =&\delta_{ll'}\hat K^{lD}\, ,\\
\int_0^{\pi}d\bar\theta (\sin\bar\theta)^{D-3}  
C_{l\,,\bar\theta}^{(D-3)/2}(\cos{\bar\theta}) C_{l'\,,\bar\theta}^{(D-3)/2}(\cos{\bar\theta}) & \nonumber\\
=\delta_{ll'}k^2\hat K^{lD} &\, , 
\end{align} 
and
\begin{equation}
\hat K^{lD}=\frac{2^{4-D}\pi\Gamma(l+D-3)}{\left(l+\frac{D-3}{2}\right)\left(\Gamma\left(\frac{D-3}{2}
\right)\right)^2\Gamma(l+1)}\,,\label{defhK}
\end{equation}
we have
\begin{equation}
K^{lD}=\hat K^{lD}{\cal A}_{D-3}\,,\label{defK}
\end{equation}
where
\begin{equation}
{\cal A}_{D-3}= \frac{2\pi^{(D-2)/2}}{\Gamma\left(\frac{D-2}{2}\right)} \, ,
\label{areasphere}
\end{equation}
is the surface of the $(D-3)$-sphere $S^{D-3}$. Note that 
$\int d\Omega^{D-2}(\cdots)={\cal A}_{D-3}\int d\bar\theta (\sin\bar\theta)^{D-3}(\cdots)$.
With the definitions (\ref{defharm}) ${\cal S}_{l\,\bar i}=-\frac{1}{k}{\cal S}_{l\,,\bar i}$,
\begin{align}
\int_0^{\pi}d\tilde\theta (\sin\bar\theta)^{D-3}
{\cal S}_l(\tilde\theta) {\cal S}_{l'}(\bar\theta)=&\delta_{ll'}{\cal A}_{D-3}^{-1}\, ,\label{intS0}
\end{align}
\begin{align}
 &\int_0^{\pi}d\tilde\theta (\sin\tilde\theta)^{D-3}\gamma^{\bar i\bar j}
{\cal S}_{l\,\bar i}{\cal S}_{l'\,\bar j} \nonumber\\
&=\int_0^{\pi}d\tilde\theta (\sin\tilde\theta)^{D-3}
{\cal S}_{l\,\bar\theta}(\bar\theta){\cal S}_{l'\,\bar\theta}(\bar\theta) \nonumber\\
&=\delta_{ll'}{\cal A}_{D-3}^{-1}\,.\label{intS}
\end{align}
Furthermore, we note that Eqs.~\eqref{eqscal3} and \eqref{defww} imply
\begin{equation}
{\cal W}_l+(D-2)\cot\bar\theta{\cal S}_{l\,,\bar\theta}+k^2{\cal S}_l=0 \,,
\end{equation} 
so that
\begin{align}
&{\cal W}_{l\,,\bar\theta}+(D-2)\cot\bar\theta{\cal S}_{l\,,\bar\theta\bar\theta}
-\frac{D-2}{\sin^2\bar\theta}{\cal S}_{l\,,\bar\theta}+k^2{\cal S}_{l\,,\bar\theta}\nonumber\\
&={\cal W}_{l\,,\bar\theta}+(D-2)\cot\bar\theta{\cal W}_{l}+(k^2-D+2){\cal S}_{l\,,\bar\theta}\,,
\end{align}
and therefore
\begin{align}
&\int_0^\pi d\bar\theta(\sin\bar\theta)^{D-3}{\cal W}_l {\cal W}_{l'} \nonumber\\
 =&\int_0^\pi d\bar\theta(\sin\bar\theta)^{D-3}{\cal W}_l\sin\bar\theta
 \left(\frac{{\cal S}_{l'\,,\bar\theta}}{\sin\bar\theta}\right)_{,\bar\theta}\nonumber\\
 =&-(D-2) \int_0^\pi d\bar\theta(\sin\bar\theta)^{D-3}{\cal W}_l\cot\bar\theta{\cal S}_{l'\,,\bar\theta}
  \nonumber\\
  &-\int_0^\pi d\bar\theta(\sin\bar\theta)^{D-3}{\cal W}_{l\,,\bar\theta}{\cal S}_{l'\,,\bar\theta}
\nonumber\\
=&(k^2-D+2)\int_0^\pi d\bar\theta(\sin\bar\theta)^{D-3}{\cal S}_{l\,,\bar\theta}{\cal S}_{l'\,,\bar\theta}\nonumber\\
=&\delta_{ll'}{\cal A}_{D-3}^{-1}\,k^2(k^2-D+2)\,.\label{intS1}
\end{align}
We thus obtain
\begin{equation}
\int d\Omega^{D-2}{\cal W}_l {\cal W}_{l'}=\delta_{ll'}k^2(k^2-D+2)\,.\label{normalK2}
\end{equation}
The perturbations $f^l_{ab}(t,r)$, $f^l_{a}(t,r)$, $H^l_L (t,r)$, $H^l_T (t,r)$ appearing in the expansion 
of the metric perturbations \eqref{mpert}
\begin{align}
h_{ab}&=f^l_{ab}{\cal S}_l(\bar\theta)\, , \\
h_{a\bar\theta}&=-\frac{1}{k} rf^l_{a}{\cal S}_l(\bar\theta)_{,\bar\theta}\,  ,\\
h_{\bar\theta\bar\theta}&=2r^2\left(H^l_L {\cal S}_l(\bar\theta)
+H^l_T \frac{D-3}{k^2(D-2)}{\cal W}_l(\bar\theta)\right)\, , \\
h_{\theta\theta}&=2r^2\sin^2\bar\theta\left(H^l_L {\cal S}_l(\bar\theta)
-H^l_T \frac{1}{k^2(D-2)}{\cal W}_l(\bar\theta)\right) \, .
\end{align}\label{mpert3}
are given by the following integrals, as follows from Eqs.~\eqref{defSC}, \eqref{defWC}, \eqref{normalK1}, \eqref{normalK2}:
\begin{align}
f^l_{ab}(t,r) =&\int d\Omega^{D-2}h_{ab}{\cal S}_l \nonumber\\
 =&\frac{{\cal A}_{D-3}}{\sqrt{K^{lD}}}\int_0^{\pi}
   d\bar\theta (\sin\bar\theta)^{D-3}h_{ab} C_l^{(D-3)/2}\, , \\
f_{a}(t,r) =&-\frac{1}{kr}\int d\Omega^{D-2}h_{a\bar\theta}{\cal S}_{l\,,\bar\theta} \nonumber\\
 =& -\frac{1}{kr}\frac{{\cal A}_{D-3}}{\sqrt{K^{lD}}}\int_0^{\pi}d\bar\theta (\sin\bar\theta)^{D-3} 
    h_{a \bar \theta }C_{l\,,\bar\theta}^{(D-3)/2}\, ,\\
H_L(t,r) =& \frac{1}{2(D-2)r^2}\int d\Omega^{D-2}\left[h_{\bar\theta\bar\theta}
           +\frac{D-3}{\sin^2\bar\theta}h_{\theta\theta}\right] {\cal S}_l \nonumber\\
         =& \frac{1}{2(D-2)r^2}\frac{{\cal A}_{D-3}}{\sqrt{K^{lD}}}\int_0^{\pi}d\bar\theta 
            (\sin\bar\theta)^{D-3} \nonumber\\
          & \times \left[h_{\bar\theta\bar\theta}+\frac{D-3}{\sin^2\bar\theta}h_{\theta\theta}\right]
            C_l^{(D-3)/2}\, ,\\
H_T(t,r) =& \frac{1}{2r^2(k^2-D+2)}\int d\Omega^{D-2} \nonumber\\
          & \times \left[h_{\bar\theta\bar\theta}-\frac{1}{\sin^2\bar\theta}h_{\theta\theta}\right]{\cal W}_l \nonumber\\
         =& \frac{1}{2r^2(k^2-D+2)}\frac{{\cal A}_{D-3}}{\sqrt{K^{lD}}}\int_0^{\pi}d\bar\theta (\sin\bar\theta)^{D-3} \nonumber\\
          & \times \left[h_{\bar\theta\bar\theta}-\frac{1}{\sin^2\bar\theta}h_{\theta\theta}\right] W_l\,,
\label{integralsfH}
\end{align}
where $h_{ab} = h_{ab}(t,r,\bar\theta)$, $h_{a\bar\theta} = h_{a\bar\theta}(t,r,\bar\theta)$, 
$h_{\bar\theta\bar\theta} = h_{\bar\theta\bar\theta}(t,r,\bar\theta)$, 
$h_{\theta\theta} = h_{\theta\theta} (t,r,\bar\theta)$, 
$C_l^{(D-3)/2} = C_l^{(D-3)/2}(\cos{\bar\theta})$ and $W_l = W_l(\cos\bar\theta)$.

We also note that the background Tangherlini metric depends on the $l=0$ harmonic only; 
the integral of its components over $l\ge2$ harmonics vanish. 
Therefore, if we decompose the space-time metric (see Appendix
\ref{app:coordtrafo}) as $g_{\mu\nu}=g^{(0)}_{\mu\nu}+h_{\mu\nu}$ with $\mu,\nu=(t,r,\bar\theta,\theta)$ and
$g^{(0)}_{\mu\nu}$ is the Tangherlini background metric, we can compute the integrals (\ref{integralsfH}) in terms of the metric
$g_{\mu\nu}$
\begin{align}
\label{eq:pertftt}
f_{tt} =& \frac{1}{\pi} \frac{{\cal A}_{D-3}}{\sqrt{K^{lD}}}
               \int d\bar{\theta}
               (\sin\bar{\theta})^{D-3} C_l^{(D-3)/2}
               \int d\theta 
               g_{tt}(\bar{\theta},\theta) \, ,\\
\label{eq:pertftr}
f_{tr} =& \frac{1}{\pi} \frac{{\cal A}_{D-3}}{\sqrt{K^{lD}}}
               \int d\bar{\theta}
               (\sin\bar{\theta})^{D-3} C_l^{(D-3)/2}
               \int d\theta 
               g_{tr}(\bar{\theta},\theta) \, , \\
\label{eq:pertfrr}
f_{rr} =& \frac{1}{\pi} \frac{{\cal A}_{D-3}}{\sqrt{K^{lD}}} 
               \int d\bar{\theta}
               (\sin\bar{\theta})^{D-3} C_l^{(D-3)/2}
               \int d\theta 
               g_{rr}(\bar{\theta},\theta) \, ,\\
\label{eq:pertfft}
f_{t} =& -\frac{1}{k r \pi} 
          \frac{{\cal A}_{D-3}}{\sqrt{K^{lD}}} \nonumber\\
       & \times \int d\bar{\theta}
                (\sin\bar{\theta})^{D-3} \partial_{\bar{\theta}} C_l^{(D-3)/2}
                \int d\theta  
                g_{t\bar{\theta}}(\bar{\theta},\theta)\, , \\
\label{eq:pertffr}
f_{r} =& -\frac{1}{ kr \pi} 
          \frac{{\cal A}_{D-3}}{\sqrt{K^{lD}}} \nonumber\\
       & \times\int d\bar{\theta}
               (\sin\bar{\theta})^{D-3} \partial_{\bar{\theta}} C_l^{(D-3)/2}
               \int d\theta  
               g_{r\bar{\theta}}(\bar{\theta},\theta)\, , \\
\label{eq:pertHL}
H_L =& \frac{1}{ 2 (D-2) r^2 \pi } 
       \frac{{\cal A}_{D-3}}{\sqrt{K^{lD}}} 
       \int d\bar{\theta}
               (\sin\bar{\theta})^{D-3} C_l^{(D-3)/2} \nonumber\\
     & \times \int d\theta 
              \left(   g_{\bar{\theta}\bar{\theta}}(\bar{\theta},\theta) 
                + (D-3) \frac{g_{\theta\theta}(\bar{\theta},\theta)}{\sin^2\bar\theta}  \right)\, ,\\
\label{eq:pertHT}
H_T =& \frac{1}{ 2 (k^2 - D + 2) r^2 \pi } 
       \frac{{\cal A}_{D-3}}{\sqrt{K^{lD}}} \nonumber\\
     & \times \int d\bar{\theta}
               (\sin\bar{\theta})^{D-3} W_l
               \int d\theta 
               \left(  g_{\bar{\theta}\bar{\theta}}(\bar{\theta},\theta) 
                - \frac{g_{\theta\theta}(\bar{\theta},\theta)}{\sin^2\bar\theta} \right) \, .
\end{align}
Furthermore, from Eqs. (\ref{eq:func1}) and (\ref{eq:pertftt}) - (\ref{eq:pertHT}) we deduce
\begin{align}
\label{eq:Ft}
F_{,t} =&  \partial_t H_L + \frac{1}{D-2} \partial_t H_T 
         + \frac{1}{k}f(r)\left( \partial_t f_r 
         + \frac{r}{k}\partial_t \partial_r H_T \right)\, , \\
\label{eq:Furt}
F^r_t =& f(r) \left( f_{rt} + \frac{r}{k} (\partial_t f_r + \partial_r f_t)
               +\frac{1}{k} f_t \right. \nonumber\\
       & \left. + \frac{2r}{k^2}(\partial_t H_T + r\partial_t\partial_r H_T)\right)  
        -\frac{r}{k}\partial_r f(r)\left( f_t + \frac{r}{k}\partial_t H_T \right)\,.
\end{align}
Conversely, since the perturbations do not depend on the $l=0$ harmonic, the background metric $g_{\mu\nu}$ can be
obtained as follows:
\begin{align}
 g_{tt}^{(0)} &=\frac{1}{K^{0D}\pi} \int_0^{\pi} d\bar{\theta} \sin^{D-3}\bar{\theta} 
  \int_0^{\pi} d \theta  g_{tt}(\bar{\theta},\theta) \,, \\
 g_{tr}^{(0)}&=0=\frac{1}{K^{0D}\pi} \int_0^{\pi} d\bar{\theta} \sin^{D-3}\bar{\theta} 
  \int_0^{\pi} d \theta  g_{tr}(\bar{\theta},\theta) \,, \\
g_{rr}^{(0)} &= \frac{1}{K^{0D}\pi} \int_0^{\pi} d\bar{\theta} \sin^{D-3}\bar{\theta} 
  \int_0^{\pi} d \theta  g_{rr}(\bar{\theta},\theta) \,.
\end{align}
Finally, to compute the areal radius $r$ we note that $g_{\bar\theta\bar\theta}=r^2+h_{\bar\theta\bar\theta}$ and
$g_{\theta\theta}=r^2\sin^2\bar\theta+h_{\theta\theta}$. Both the perturbations $h_{\bar\theta\bar\theta}$ and
$h_{\theta\theta}$ contain harmonics of different type (${\cal S}_l$, ${\cal S}_{l\,,\bar i\bar j}$); to extract the
background we need the combination in Eq.~\eqref{eq:pertHL}:
\begin{align}
r^2 =& \frac{1}{(D-2)K^{0D}\pi} \int_0^{\pi} d\bar{\theta} \sin^{D-3}\bar{\theta} \int_0^\pi d\theta \nonumber\\
     & \times \left[g_{\bar\theta\bar\theta}+(D-3)\frac{g_{\theta\theta}}{\sin^2\bar\theta}\right]\,.\label{reparam}
\end{align}


\bibliographystyle{h-physrev4}
\bibliography{collision}

\end{document}